\pdfoutput=1
\documentclass[acmsmall]{acmart}
\AtBeginDocument{%
  \providecommand\BibTeX{{%
    Bib\TeX}}}

\setcopyright{acmlicensed}
\acmJournal{TOSEM}
\acmYear{2024} 
\acmVolume{1} 
\acmNumber{1} 
\acmArticle{1} 
\acmMonth{1}
\acmDOI{10.1145/3699597}

\usepackage{amsmath,amssymb,amsfonts}
\usepackage{algorithmic}
\usepackage{graphicx}
\usepackage{textcomp}
\usepackage{xcolor}
\usepackage{balance} 
\usepackage{url} 
\def\BibTeX{{\rm B\kern-.05em{\sc i\kern-.025em b}\kern-.08em
    T\kern-.1667em\lower.7ex\hbox{E}\kern-.125emX}}
\usepackage{textcomp}
\usepackage{booktabs}
\usepackage{enumitem}
\usepackage{multirow}
\usepackage{subfig}
\usepackage{listings}
\usepackage{color}
\usepackage{threeparttable}
\usepackage{diagbox}
\usepackage{graphicx}
\usepackage{verbatim}
\usepackage{url}
\usepackage{soul}
\soulregister\cite7
\soulregister\ref7
\usepackage{graphicx}
\usepackage{multirow}
\usepackage{hhline}
\usepackage{comment}
\usepackage{float}
\usepackage{array,booktabs}
\usepackage{marvosym} 
\usepackage{pifont}
\usepackage{tcolorbox}
\usepackage{bm}
\usepackage[ruled,linesnumbered]{algorithm2e}
\usepackage{amsthm}

\newcommand{\appname}{{\sc SmartBT}\xspace}
\newcommand{\appnamebold}{{\sc \textbf{SmartBT}}\xspace}
\newcommand{\find}[1]{
\begin{tcolorbox}[leftrule=1mm,toprule=0mm,bottomrule=0mm,left=1pt,right=2pt,top=2pt,bottom=2pt] 
#1
\end{tcolorbox}
}

\begin{document}

\title[Automating Comment Generation for Smart Contract from Bytecode]{Automating Comment Generation for Smart Contract \\ from Bytecode}

\author{Jianhang Xiang}
\affiliation{%
  \institution{The State Key Laboratory of Blockchain and Data Security, Zhejiang University}
  \city{Hangzhou,}
  \country{China}
}
\author{Zhipeng Gao}
\authornote{This is the corresponding author}
\affiliation{%
  \institution{The State Key Laboratory of Blockchain and Data Security, Zhejiang University}
  \city{Shanghai,}
  \country{China}
}
\author{Lingfeng Bao}
\affiliation{%
  \institution{The State Key Laboratory of Blockchain and Data Security, Zhejiang University}
  \city{Hangzhou,}
  \country{China}
}
\affiliation{%
  \institution{Hangzhou High-Tech Zone (Binjiang) Blockchain and Data Security Research Institute}
  \city{Hangzhou,}
  \country{China}
}
\author{Xing Hu}
\affiliation{%
  \institution{The State Key Laboratory of Blockchain and Data Security, Zhejiang University}
  \city{Hangzhou,}
  \country{China}
}
\author{Jiayuan Chen}
\affiliation{%
  \institution{The Ohio State Univeristy}
  \city{Columbus}
  \country{United States}
}
\author{Xin Xia}
\affiliation{%
  \institution{The State Key Laboratory of Blockchain and Data Security, Zhejiang University}
  \city{Hangzhou,}
  \country{China}
}
\renewcommand{\shortauthors}{Xiang et al.}

\begin{abstract}
Recently, smart contracts have played a vital role in automatic financial and business transactions. 
To help end users without programming background to better understand the logic of smart contracts, previous studies have proposed models for automatically translating smart contract source code into their corresponding code summaries. 
However, in practice, only 13\% of smart contracts deployed on the Ethereum blockchain are associated with source code. 
The practical usage of these existing tools is significantly restricted. 
Considering that bytecode is always necessary when deploying smart contracts, in this paper, we first introduce the task of automatically generating smart contract code summaries from bytecode. 
We propose a novel approach, named {\sc \textbf{SmartBT}} (\underline{\textbf{Smart}} contract \underline{\textbf{B}}ytecode \underline{\textbf{T}}ranslator) for automatically translating smart contract bytecode into fine-grained natural language description directly.  
Two key challenges are posed for this task: structural code logic hidden in bytecode and the huge semantic gap between bytecode and natural language descriptions. 
To address the first challenge, we transform bytecode into CFG (Control-Flow Graph) to learn code structural and logic details. 
Regarding the second challenge, we introduce an information retrieval component to fetch similar comments for filling the semantic gap. 
Then the structural input and semantic input are used to build an attentional sequence-to-sequence neural network model. 
The copy mechanism is employed to copy rare words directly from similar comments and the coverage mechanism is employed to eliminate repetitive outputs.  
The automatic evaluation results show that {\sc SmartBT} outperforms a set of baselines by a large margin, 
and the human evaluation results show the effectiveness and potential of {\sc SmartBT} in producing meaningful and accurate comments for smart contract code from bytecode directly. 
\end{abstract}




\maketitle
\section{Introduction}\label{sec:intro}

Smart contract, which was created by Nick Szabo~\cite{szabo1997formalizing}, is a program (mainly written by Solidity) that can be triggered and executed on the Ethereum Blockchain. 
The Ethereum blockchain has experienced remarkable growth along the blockchain technology, the average amount of smart contracts deployed on Ethereum has exceeded 4,200 each month, and the average number of active Ethereum addresses has exceeded 82 million per month. 
As a result, Ethereum has become one of the largest cryptocurrency platforms (with an overall market capitalization surpassing 228.38 billion in USD) and smart contracts are increasingly used to automate financial and business transactions. 

Despite the great success of smart contracts, notable concerns have also emerged, especially for novice users who are unfamiliar with smart contracts. 
Among them, \textbf{false advertising} stands out as a crucial issue faced by smart contract end users. 
For instance, smart contract end users may experience ICO (Initial Coin Offer) scams when investing in cryptocurrencies. 
The scammers make false claims about the project's innovative technology and/or business logic, enticing investors with promises of substantial returns on investment. 
However, the smart contracts they deployed lack such actual utility, value, or even existence. 
The mismatch between the smart contract code implementations and 
the content described in their white papers, websites, and announcements has led to significant financial losses for investors. 
As pointed out by ~\cite{Bitcoinafrica}, ten of the most high-profile ICO scams swindled 687.4 million USD from unsuspecting investors.
To help end users find the inconsistency between the smart contract source code and their documentation, researchers have investigated the ways of automatically translating Solidity source code into fine-grained English descriptions~\cite{hu2021automating,shi2023machine,yang2022ccgir}. 
For example, Hu et al.~\cite{hu2021automating} introduced SmartDoc, a deep-learning-based model to generate user notice for smart contracts. 
In this way, end users without programming knowledge can understand and learn the logic of original smart contracts.
Shi et al.~\cite{shi2023machine} presented a reinforcement learning model, named SolcTrans, to generate comments of the Solidity source code via AST traversal and Probalistic Context-Free Grammar rules. 
\textbf{All of the previous studies focus on generating high-level summaries from Solidity source code, however, it is not guaranteed that the Solidity source code of every smart contract can be obtained. 
}

Typically, when developers deploy smart contracts on the Ethereum blockchain, they first compile the contract source code to bytecode and then the bytecode is deployed to the blockchain. 
Developers can choose to upload their source code for verification but this is optional. 
According to statistics obtained by Zellic~\cite{Zellic}, \textbf{they collected more than 30 million smart contracts deployed on the Ethereum blockchain, only 13\% of them are associated with source code.} 
In other words, the existing methods which rely on the availability of Solidity source code, are incapable of handling the majority of smart contracts on the Ethereum blockchain where source code are missing. 
As a result, the practical usage of these tools is significantly limited and severely restricted. 

Although the source code of the smart contracts may not be publicly accessible, the bytecode of every smart contract is available upon deployment on the blockchain.
\textbf{In this study, we first investigate the possibility of generating smart contract code summaries directly from bytecode, bypassing the existing methods that rely on the original source code.} 
However, generating comments from bytecode is a difficult task with respect to the following challenges:
\begin{enumerate}
     \item \textit{Learning structural information from bytecode.} 
     The Solidity bytecode is generated by compiling Solidity source code into an instruction set that can be read and executed by the Ethereum Virtual Machine (EVM). 
     Unlike human-readable source code, the bytecode is a series of hexadecimal numbers, which are not easily understandable by individuals without specialized knowledge. 
     Therefore, how to extract the information about the contract's structure and behavior as well as learning the logic and functionality pattern encoded within Solidity bytecode presents a significant challenge.  
     \item \textit{Semantic gap between bytecode and comments.} 
     The semantic gap refers to the disparity between the low-level EVM bytecode and the high-level explanations conveyed by human-written comments. 
     Compared with source code, the semantic gap between bytecode and natural language comment is even more larger, how to properly fill the gap and effectively transform bytecode into meaningful natural language comments is another challenge for this study. 
\end{enumerate}

In this work, to help end users better understand the logic details of the smart contracts deployed on the Ethereum blockchain, we propose a novel neural network model, named \appname (\underline{Smart} contract \underline{B}ytecode \underline{T}ranslator), which can automatically translate smart contract bytecode into human-readable code summaries. 
The generated code descriptions can be used as function code comments to help users understand and participate in interacting with smart contracts more easily and safely. 
In particular, we first collect 30,742 $\langle bytecode, comment \rangle$ pairs from 54,739 deployed smart contracts. 
To extract structural information from the bytecode, we convert each smart contract bytecode into a Control-Flow Graph (CFG) to learn the contract structure from the encoded bytecode. 
To fill the semantic gap between bytecode and comment, we introduce the IR (Information-retrieval) augmented module to fetch relevant comments from contracts with similar CFGs. 
Finally, we build an IR-augmented sequence-to-sequence model by incorporating the structural input (i.e., CFGs) and semantic input (i.e., relevant comments). 
Moreover, we have introduced the \textit{copy} mechanism to copy rare words from the IR module and the \textit{coverage} mechanism to eliminate the word repetition problem. 
The automatic and human evaluation results show the advantage and superiority of {\sc SmartBT} for generating comments from bytecode for smart contracts. 
The paper makes the following contributions:  
\begin{itemize}
    \item The existing comment generation methods rely on the availability of source code, while 90\% deployed smart contracts' source code are missing on the Ethereum blockchain. 
    In this study, we first proposed the new task of generating smart contract comments from the bytecode level. 
    \item We build the first dataset for the smart contract bytecode comment generation task. 
    In particular, we have collected 30,742 $\langle bytecode, comment \rangle$ pairs of different functions from 54,739 verified smart contracts. 
    Each function bytecode is converted into an intermediate representation of CFG to provide useful structural information for downstream tasks. 
    \item To the best of our knowledge, {\sc SmartBT} is the first model to investigate the possibility of generating smart contract comments directly from the bytecode. 
    We introduce the CFG and IR augmented components to fill the gap between the bytecode and natural language comments. 
    We extensively evaluate the {\sc SmartBT} on real-world deployed smart contracts, {\sc SmartBT} is shown to outperform several baselines and reduce the user's efforts in understanding smart contracts. 
    \item We have released a replication package~\cite{smartbt_dataset}, including the dataset and source code of {\sc SmartBT}, to facilitate other researchers and practitioners to repeat our work and verify their own ideas. 
\end{itemize}

The rest of the paper is organized as follows. Section~\ref{background} presents the background of our research. 
Section~\ref{approch} presents our approach details.  Section~\ref{setup} presents the experiment setup.  Section~\ref{results} presents our research questions and experimental results.  
Section~\ref{threats} presents the threats to validity. Section~\ref{relatedwork} presents related works. 
Finally, Section~\ref{conclusion} presents the conclusion and future work.  

\section{Background}
\label{background}

\subsection{Smart Contracts}\label{subsec:contract}
\textit{Smart contract}, a term coined by Nick Szabo in 1994~\cite{szabo1996smart}, was proposed as a computerized transaction protocol that executes the contractual terms of an agreement. 
Recently, along with development of the blockchain technology, smart contracts can be essentially implemented on top of blockchains (i.e., Ethereum), which are referred to as code scripts to execute certain tasks once predefined conditions are met.

\begin{figure} 
  \centering
  \includegraphics[width=0.85\linewidth]{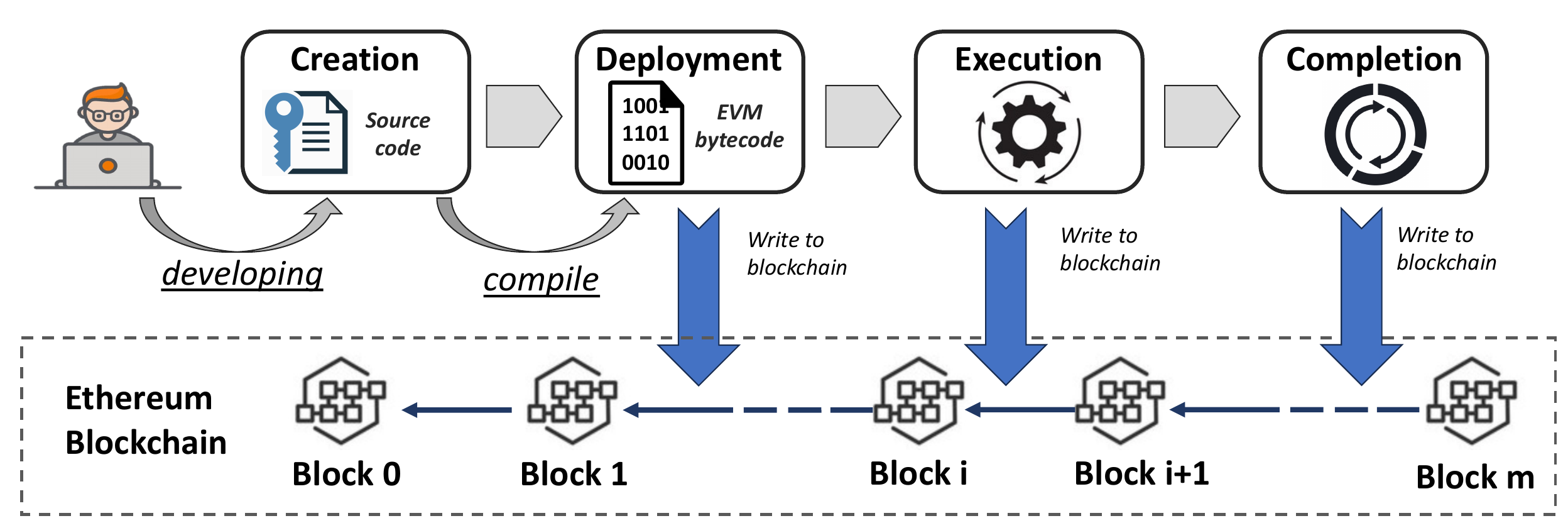} 
  \caption{Life Cycle of Smart Contracts} 
  \label{fig:bg}
\end{figure}

Particularly, smart contracts are programs that run on the Ethereum blockchain. 
The contractual clauses are converted into executable computer programs. 
The logical connections between contractual clauses are also been preserved in the form of logical flows in programs (e.g., the \texttt{if-else} statement). 
If any condition in a smart contract is satisfied, the triggered statement will automatically execute the corresponding function in a predictable manner. 
The execution of each contract statement is then recorded as an immutable transaction stored in the blockchain system. 
For example, Bob and Alice have an agreement on the penalty of violating the contract. 
If Bob breaches the contract, the corresponding penalty (as specified in the contract) will be automatically deducted from Bob's deposit.

As shown in Fig.~\ref{fig:bg}, the life cycle of smart contracts consists of four consecutive phases:
1) \textit{Smart Contract Creation:} Several involved parties (e.g., stakeholders, lawyers) will first reach a contractual agreement after discussions. 
Then software developers convert this agreement written in natural languages into smart contract(s) written in computer languages (e.g., Solidity, Vyper). 
2) \textit{Smart Contract Deployment:} The validated smart contracts then can be deployed on top of the Ethereum blockchain. 
Since Ethereum uses EVM (Ethereum Virtual Machine) to execute smart contracts, to deploy a smart contract on Ethereum, the contract source code (e.g., Solidity) needs to be compiled into EVM bytecode and the bytecode will be stored on the blockchain. 
3) \textit{Smart Contract Execution:} After deployment, the smart contracts are triggered by events. 
These events can be external (e.g., payment received) or internal (e.g., specific date or time). 
Once an event satisfies the predefined conditions, the corresponding statements will be automatically executed. A transaction is executed and validated by miners in the Ethereum blockchain. 
4) \textit{Smart Contract Completion:} After execution, the transactions during the execution, as well as the updated states, are permanently stored in blockchains. 
Meanwhile, the digital assets have been transferred from one party to another (e.g., money transfer from the buyer to the supplier). 
Notably, only EVM bytecode is deployed and stored on the blockchain, while the source code remains off-chain and inaccessible.

From the developers' perspective, the developer first needs to write source code for smart contracts with programming languages (e.g., Solidity). 
Then the developer uses compilers (e.g., \texttt{Solc}) to convert source code into EVM bytecode. 
The developer deploys the bytecode to the Ethereum blockchain and saves it at an address. 
Particularly, two types of EVM bytecode are associated with deployment, i.e., \textit{creation code} and \textit{runtime code}. 
\textit{Creation code} is responsible for the creation of the contract (e.g., setting up the constructor and initializing constructor variables), which is executed only once during deployment. 
Unlike \textit{creation code}, \textit{runtime code} doesn't include the constructor details. 
Once the \textit{creation code} is executed, \textit{runtime code} is stored on-chain for anyone to interact with. 
\textit{Runtime code} describes the contract, any on-chain interaction with the smart contract means an interaction with the \textit{runtime code}. 
In this work, we refer to smart contract \textit{runtime code} as bytecode for short in the rest of this paper.

\subsection{Neural Machine Translation}
Neural Machine Translation (NMT) is an end-to-end learning framework for automated translation. 
It is a deep learning-based approach and has made rapid progress in recent years~\cite{hu2018deep, gao2020generating, hu2021automating, gao2023know}. 
NMT has shown impressive results surpassing those of phrase-based or rule-based systems while addressing shortcomings such as the need for manually crafted features. 
NMT models usually consist of an encoder-decoder structure. 
The encoder encodes the input sequence into a fixed-length vector, which represents the semantic and contextual information of the source language sentence. 
The decoder gradually generates translated output sequences based on the vectors encoded by the encoder and the previously generated target language parts. 


NMT has proven to be effective in bridging the gap between different languages in natural language processing. 
The NMT framework has also been successfully applied to various software engineering tasks~\cite{qiu2021deep, hu2021automating, gao2020generating, gao2020technical}, including comment generation~\cite{hu2021automating}. 
Software engineering researchers view comment generation as a variant of the machine translation problem between source code and natural language. 
However, all previous studies focus on generating comments from source code. 
In this study, we first explore the possibility of whether the NMT framework can be applied to comment generation from Solidity bytecode. 


\section{Motivation}
\label{moti}

\begin{figure} 
  \centering
  \includegraphics[width=0.9\linewidth]{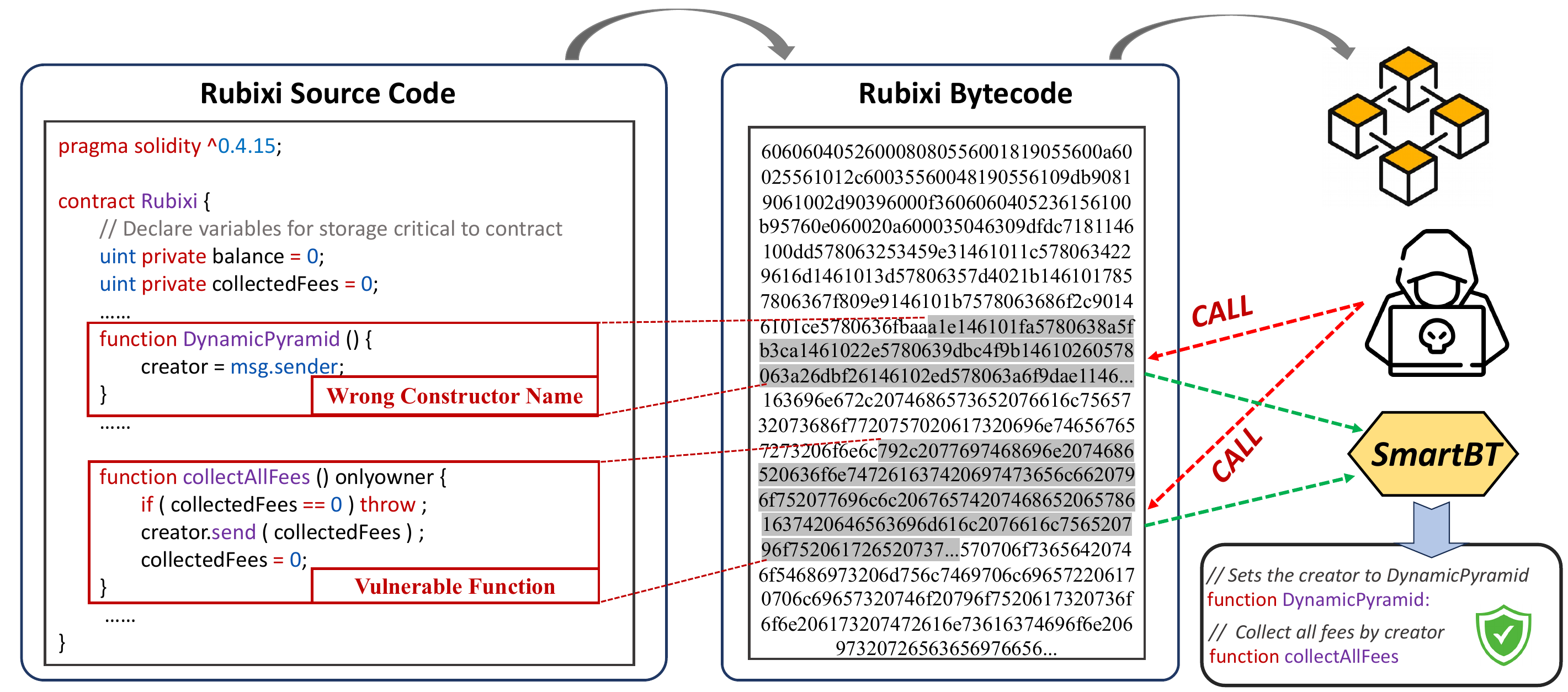} 
  \caption{Motivating Example of Using \appnamebold} 
  \label{fig:moti}
\end{figure}

In this section, we provide a motivating example to explain the usage of our tool in practice. 
Fig.~\ref{fig:moti} shows a common vulnerability, namely \textit{wrong constructor name}, in smart contracts. 
Constructors are special functions that are called only once during the contract creation. 
They often perform critical, privileged actions such as setting the owner of the contract. 
Constructor names have to be the same as the contract class that contained it. 
If the constructor method is inconsistent with the contract class (a.k.a., missnamed), security issues will be introduced.

Fig.~\ref{fig:moti} demonstrates a real smart contract with the \textit{wrong constructor name} weakness. 
The smart contract \texttt{Rubixi} uses \texttt{DynamicPyramid} instead of \texttt{Rubixi} as a constructor. 
Because of this inconsistency, the contract did not assign the owner upon contract creation. 
As a result, anyone who calls this \texttt{DynamicPyramid()} function can assign themselves as the owner of the contract.
After granting the ownership, then they can easily collect contract fees generated by participating players (e.g., calling function \texttt{collectAllFees()}). 
With our tool, even if contract \texttt{Rubixi} is not open-sourced, \textsc{SmartBT} can successfully generate a comment, i.e., \textit{sets the creator to DynamicPyramid}, for the  \texttt{DynamicPyramid()} function from its bytecode. 
It can help users notice this inconsistency between contract name (\texttt{Rubixi}) and constructor name (\texttt{DynamicPyramid}), and better make informed decisions.

\section{Approach}
\begin{figure*} 
  \centering
  \includegraphics[width=0.95\linewidth]{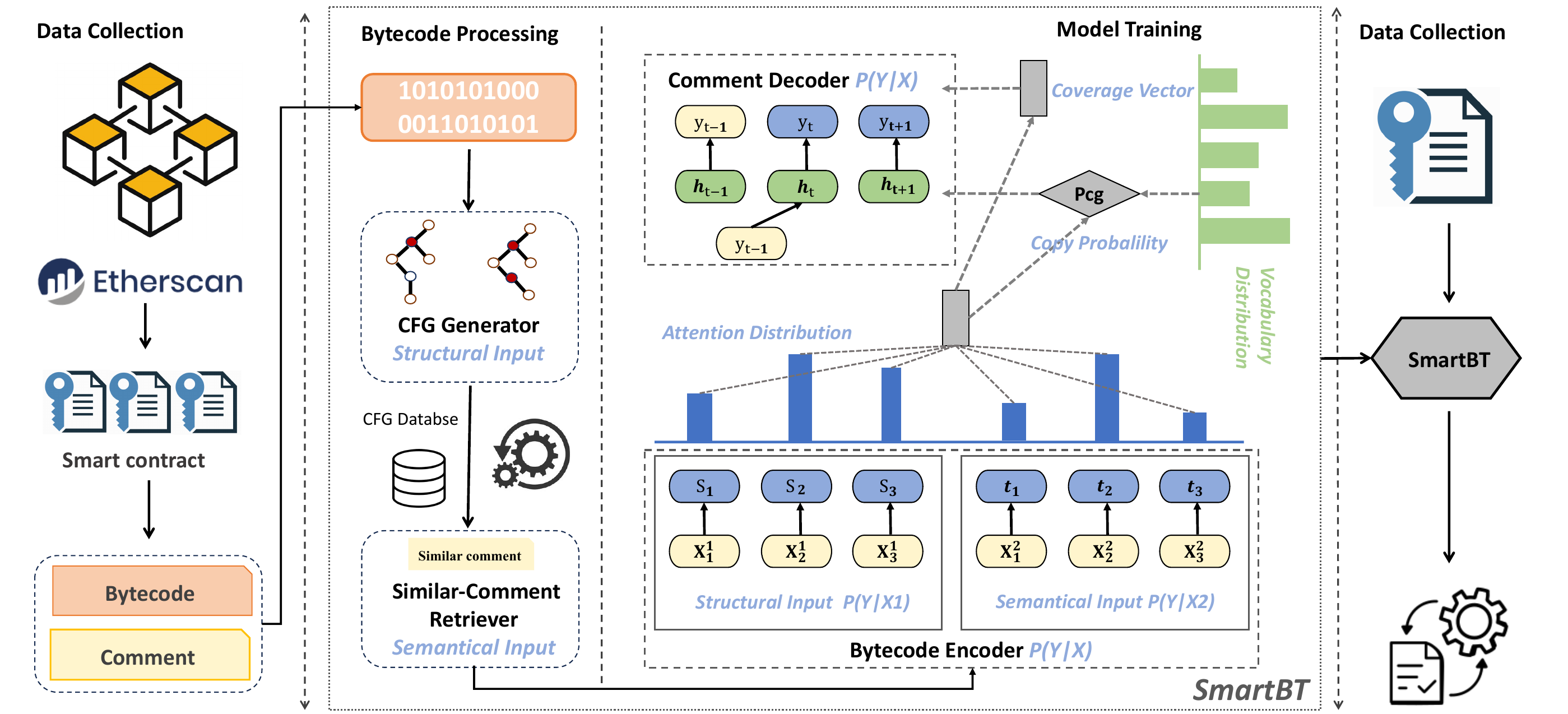} 
  \caption{Workflow of \appnamebold} 
  \label{fig:workflow}
\end{figure*}

In this section, we first define the task of comment generation from bytecode, then present the details of our approach. 
Figure~\ref{fig:workflow} demonstrates the workflow of the {\sc SmartBT}. 
Our approach is primarily composed of three stages: bytecode preprocessing, model training, and comment generation.

\subsection{Task Definition}
The main purpose of our work is to improve the understanding of a smart contract from its bytecode directly. 
We thus propose a novel task in this paper - generating smart contract code comments from their corresponding bytecode. 
Inspired by the great success of using the NMT framework in previous studies~\cite{hu2021automating}, We formulate our task as a sequence-to-sequence learning problem.  


Given that $X$ is the input bytecode sequence of a smart contract function, our target is to generate its corresponding comment $Y$ describing the function. 
In particular, our objective is to learn the underlying conditional probability distribution ${P_{\theta}}(Y|X)$. 
In other words, the goal is to train a model $\theta$ using $\langle X, Y \rangle$ pairs such that the probability ${P_{\theta}}(Y|X)$ is maximized over the given training dataset. The training objective function can be formulated as maximizing the log-likelihood:
\begin{equation}\label{formula1}
{\theta}^* = \arg\max_{\theta} \sum_{\langle X, Y \rangle} \log P_{\theta}(Y|X).
\end{equation}

\label{approch}


\subsection{Bytecode Processing}


Because we directly deal with smart contracts bytecode, two key challenges are posed for this study. 
\begin{itemize}
    \item \textbf{Challenge 1:} How can we effectively capture the structural information and logic details encoded in smart contract bytecode? 
    \item \textbf{Challenge 2:} How can we effectively bridge the gap between semantic gap between the bytecode and natural language comments? 
\end{itemize}
To address these two key challenges, we process the smart contract bytecode with two components: a \textbf{CFG Generator} and a \textbf{Similar-Comment Retriever}. 
\textbf{CFG Generator} is responsible for constructing input containing structural information, \textbf{Similar-Comment Retriever} is responsible for constructing input associated with semantic information,  then these two types of inputs are used to train an end-to-end model for automatically generating comments from contract bytecode.

\subsubsection{\textbf{CFG Generator}}
To deploy a smart contract to Ethereum, its source code needs to be compiled
to bytecode and stored on the blockchain. 
There are 140 unique opcodes by April 2019~\cite{wood2014ethereum}, and each opcode is represented by a hexadecimal number. 
EVM (Ethereum Virtual Machine) uses these opcodes to execute the task. 
When a transaction needs to be executed, EVM will first split the bytecode into bytes. Each byte represents a unique instruction called opcode. 
For example, for the bytecode \texttt{0x6070604001}, EVM first splits the bytecode into bytes (i.e., \texttt{0x60}, \texttt{0x70}, \texttt{0x60}, \texttt{0x40}, \texttt{0x01}). 
EVM then executes the first byte \texttt{0x60} which refers to the opcode \texttt{PUSH1}. 
\texttt{PUSH1} pushes the one byte data (\texttt{0x70}) to the EVM stack. Then EVM executes the third byte \texttt{0x60} and pushes \texttt{0x40} into the stack. 
Finally, EVM executes \texttt{0x01} which refers to opcode \texttt{ADD}. 
\texttt{ADD} obtains two values from the stack (i.e., \texttt{0x70} and \texttt{0x40}) and perform the sum operation. 
As can be seen, the smart contract bytecode contains detailed contract execution logic and structural information. 
To capture the structural information and detailed logic, we convert the smart contract bytecode into CFG (Control-Flow Graph) representations. 
A CFG is a graphical representation of the code control flow during the execution of a contract. 
Each node in the CFG represents a basic block, which contains a straight-line piece of opcode without any jumps or jump targets. 
Blocks are connected by edges which represent jumps to form a control flow graph. 
The CFG representations can capture all the possible flows of execution of all code blocks and can reflect the real-time execution of a code fragment, which is valuable for extracting structural information encoded within the contract bytecode.  

\begin{figure} 
  \centering
  \includegraphics[width=0.8\linewidth]{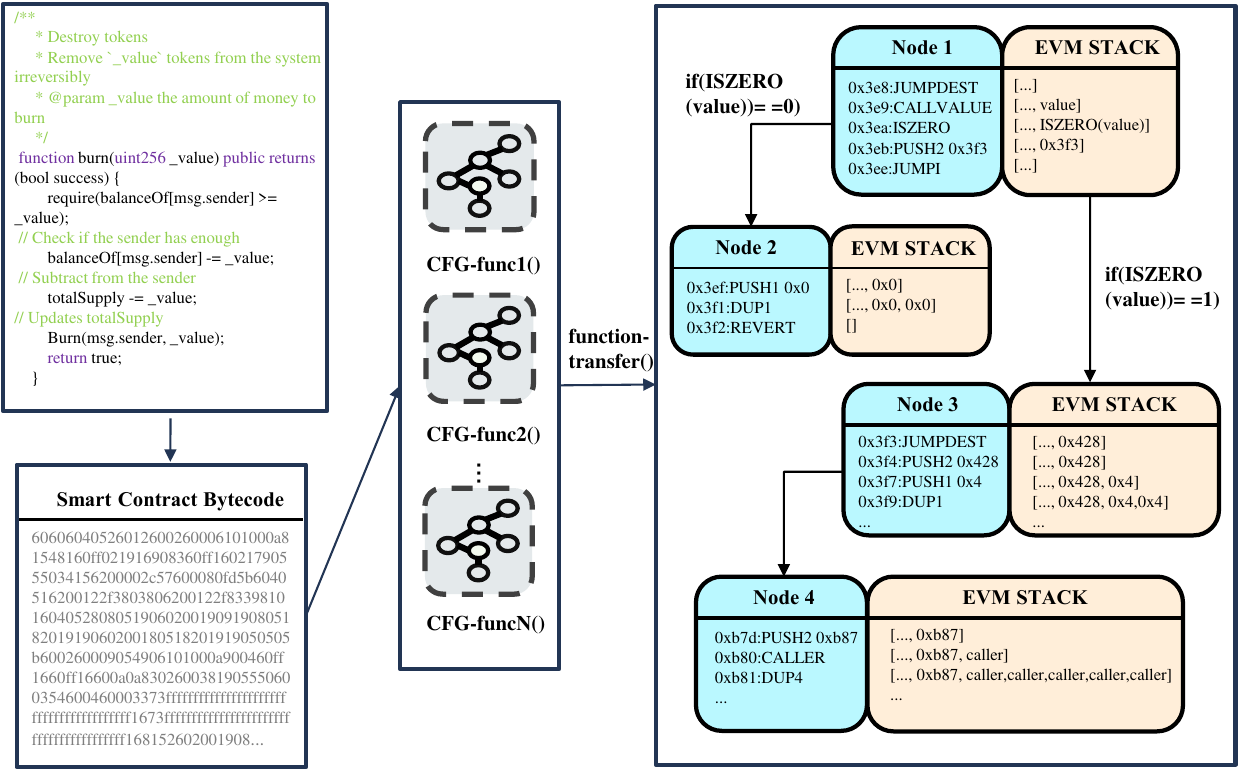} 
  \caption{Example of Solidity Source Code, Bytecode and CFG} 
  \label{fig:cfg}
\end{figure}

For a given smart contract bytecode, we first extract the CFG for each function within the smart contract. 
As shown in Figure~\ref{fig:cfg}, each smart contract may contain a number of functions, after the extraction process, each function is converted into a CFG. 
In this work, we use the tool \texttt{evm\_cfg\_builder} for CFG extraction, the tool can reliably recover a CFG from EVM bytecode using a dedicated value set analysis. 
However, the CFG cannot be directly fed into a sequence model, we further traverse the CFG by utilizing DFS (Depth-First-Search) and generate the target CFG sequence. 
In particular, we first identify the root node of each CFG, we then serialize CFG into a sequence by using DFS algorithms. 
The symbol \texttt{->} is used when traversing from one block to another, maintaining the structural information of different blocks. 
So far, we have converted each function of the bytecode into a CFG representation by our \textbf{CFG Generator}, and the structural inputs are prepared for the downstream model training. 



\subsubsection{\textbf{Similar-comment Retriever}} 
Compared with source code, the semantic gap between bytecode and natural language code descriptions is even larger, which poses another key challenge for translating bytecode into code summaries. 
To bridge the semantic gap between bytecode and comments, we introduced another component, namely \textbf{Similar-comment Retriever}. 
This component is responsible for retrieving similar comments from our codebase and using them as semantic inputs for our model. 
Our goal is to leverage these retrieved similar comments as valuable references for improving our translation process.

Specifically, we have stored all the functions' CFG and their corresponding comments as our codebase. 
When presented with the bytecode of a new function, we initially convert it into its CFG using the \textbf{CFG Generator}. 
Subsequently, we employ an information retrieval algorithm, specifically BM25~\cite{robertson2009probabilistic} for this study, to query our codebase and search for similar functions by comparing similarities between CFGs.
The top similar functions are returned and their corresponding comments are extracted as our semantic inputs. 
The \textbf{Similar-comment Retriever} component acts as a critical link in our model, providing a way to connect bytecode with semantic relevant natural language descriptions, which can enhance the final inference of our approach when generating code summaries. 
So far, the semantic inputs have also been prepared for our model building. 

\subsection{Bytecode Encoder} 
After the structural inputs, i.e., CFGs, and semantic inputs, i.e., similar-comments, are prepared, we concatenate these two inputs and feed them sequentially into our Bytecode Encoder. 
We add a special token \texttt{[SEP]} between semantic input and structural input to further separate natural language (i.e., comments) and code implementations (i.e., CFGs), which has been proven to be effective for bridging the gap between heterogeneous data~\cite{wang2021codet5, gu2020codebert}. 

The Bytecode Encoder uses Bidirectional Recurrent Neural Networks (BRNN), which can improve the model's ability to understand the input sequence from both forward and backward directions, capture more comprehensive context information, and help the decoder generate more accurate target sequences. 
The Bytecode Encoder aims to learn contextual representations from bytecode processing outputs (including generated CFGs and retrieved similar-comments). 
The concatenated inputs, including the smart contract CFG and its associated similar comments, are embedded into a vector before being fed into the Bytecode Encoder. 
The calculation equation of forward RNN and reverse RNN are:
\begin{equation}\label{for:brnn1}
h_t^f=\operatorname{BRNN}^f\left(x_t, h_{t-1}^f\right),
\end{equation}
\begin{equation}\label{for:brnn2}
h_t^b=\operatorname{BRNN}^b\left(x_t, h_{t+1}^b\right).
\end{equation}
Here $h_t^{f}$ represents the hidden state of forward RNN at time step $t$, $h_t^{b}$ represents the hidden state of reverse RNN at time step $t$, and $x_t$ represents the input sequence at time step $t$. 
The hidden states of forward RNN and reverse RNN can be calculated by the order of time steps, or by reverse calculation (starting from the last time step). 
In BRNN, the hidden states of these two directions can be concatenated or merged to obtain a complete bidirectional representation. 
Here, we mainly take the concatenation operation. As shown in Eq.~\eqref{for:brnn3}, $h_t^f$ is the forward hidden state, $h_t^b$ is the reverse hidden state, and $h_t$ is the hidden state after splicing. 
Here $[;]$ represents the splicing operation. 
By concatenation, BRNN provides richer bidirectional context information for subsequent decoding.
\begin{equation}\label{for:brnn3}
h_t=\left[h_t^f ; h_t^b\right].
\end{equation}
\subsection{Comment Decoder}
{\sc SmartBT} uses a 2-layer LSTM network as its decoder, which refers to a specific recurrent neural network structure that consists of two LSTM layers stacked together. 
Each LSTM layer has its own set of LSTM cells responsible for capturing and storing information over time. The output of one LSTM layer serves as the input to the next layer, enabling the network to learn hierarchical representations of sequential data. 
This deeper architecture enables the model to capture more complex dependencies and make more complex predictions than single-layer LSTM. 
For the decoding process, we first have the following assumptions: $t$ is the time step, and $h_t$ is the hidden states at $t$ time step. 
In our task, each token of the contract function comment will be embedded into a vector, and we assume that $Cword_t$ is the target word at $t$ of the ground truth comments, $y_t$ is the embedding vector of $Cword_t$. 
For the first and second layers of LSTM, the calculation process is as follows: 
\begin{equation}\label{for:lstm1}
{L}1_t=\operatorname{LSTM}_1\left(y_t, {h}_{t-1}\right),
\end{equation}
\begin{equation}\label{for:lstm2}
{L}2_t=\operatorname{LSTM}_2\left(y_t, \mathbf{L}1_{t}\right).
\end{equation}
During training, at each time step $t$, the first layer LSTM takes the embedding vector $y_t$ of the target word $Cword_t$ and the previous state $h_{t-1}$ as input, and concatenates them to produce the output hidden state of the first layer. 
For the second layer of LSTM, the initial value of $h_t$ is the output hidden state of the first layer of LSTM. 
The decoder produces one symbol at a time and stops when the \texttt{END} symbol is emitted. 
The only change with the decoder at the testing time is that it uses the output from the previous word emitted by the decoder in place of $Cword_{t-1}$, since there is no access to a ground truth then.

\subsection{Incorporating Attention Mechanism}
In a traditional NMT framework, the encoder processes the input sequence and encodes it into a fixed-size context vector. 
This fixed-size context vector can become a bottleneck when dealing with long sequences or capturing important information from different parts of the input. 
By using Bahdanau Attention~\cite{bahdanau2014neural} as the global attention mechanism, our \textbf{Comment Decoder} can focus on different parts of the input sequence dynamically. 
In particular, We model the attention ~\cite{bahdanau2014neural} distribution over words in the source input sequence. 
We calculate the attention $(a^{t}_{i})$ over the $i^{th}$ input sequence token as:

\begin{equation}\label{equ:attention}
    e^{t}_{i} = v^{t}\textnormal{tanh}\left(W_{eh}h_{i} + W_{sh}s_{t} + b_{att} \right)
\end{equation}
\begin{equation}
     a^{t}_{i} = \textnormal{softmax} \left( e^{t}_{i} \right)
\end{equation}
Here, $v^{t}$, $W_{sh}$ and $b_{att}$ are model parameters to be learned, and $h_{i}$ is the concatenation of forward and backward hidden states of our \textbf{Bytecode Encoder}. We use this attention $a^{t}_{i}$ to generate the context vector $c^{*}_{t}$ as the weighted sum of encoder hidden states :

\begin{equation}
\mathbf{c}^{*}_{t} = \sum_{i=1,..,|\mathbf{x}|} a^{t}_{i} \mathbf{h}_i
\end{equation}

We further use the $c^{*}_{t}$ vector to obtain a probability distribution over the words in the vocabulary as follows:
\begin{equation}
P = \textnormal{softmax} \left(\mathbf{W}_{v}[s_{t}, c^{*}_{t}] + b_{v} \right)
\end{equation}

where $W_{v}$ and $b_{v}$ are model parameters. 
Thus during decoding, the probability of generating a target word is $P(Cword)$. 
During the training process for each word at each timestamp, the loss associated with the generated comment is:

\begin{equation}
Loss = -\frac{1}{T} \sum^{T}_{t=0}logP(Cword_{t})
\end{equation}

The \textit{attention} mechanism allows the model to focus on the most relevant parts of the input sequence as needed. 
Instead of relying solely on the last hidden state, the \textit{attention} mechanism allows our \textbf{Comment Decoder} to consider all the hidden states from the \textbf{Bytecode Encoder}. 
It assigns different attention weights to each hidden state, indicating its relevance to the current decoding step. 
This ability to amplify the signal from the CFG input and similar comments input makes attention models produce better results than models without attention.

\subsection{Incorporating Copy Mechanism}
The \textit{copy} mechanism~\cite{gu2016incorporating} is commonly used in tasks such as machine translation and text summarization. 
It is used to facilitate the model to copy tokens from the input sequence to the generated output sequence. 
This is because some words are much less frequent than others, thus it is highly unlikely for a decoder that is solely based on a language model to generate such a word with very rare occurrences in
a corpus. 
In such cases, the possibly rare words in the input sequence might be required to be \textit{copied} from our input sequence to the target comment. 
Therefore, we incorporate a \textit{copy} mechanism to handle such rare words problem for our comment generation tasks.

Specifically, we calculate $p_{cg} \in [0,1]$, which determines whether to generate a word from the vocabulary or to copy the word directly from the input sequence, based on our previous attention distribution $a_{i}^t$: 
\begin{equation}
    p_{cg} = sigmoid(W^{T}_{eh}c^{*}_{t} + W^{T}_{sh}s_{t} + W_{x}x_{t} + b_{cg})
\end{equation}
Here $W_{eh}$, $W_{sh}$, $W_{x}$ and $b_{cg}$ are trainable model parameters. 
The final probability of decoding a word is specified by the mixture model:
\begin{equation}\label{for:copy2}
{P}^{*}(Cword)=p_{cg}\cdot \sum_{}^{} a_i^t+(1-p_{cg})\cdot p(Cword).
\end{equation}
Where ${P}^{*}(Cword)$ is the final distribution over the union of the vocabulary and the input sequence.
For a word not in our output vocabulary, the probability will be ${P}^{*}(Cword)$ = 0, and in such cases, our model will replace the $\textbf{<unk>}$ token for out-of-vocabulary words with a word in the input sequence having the highest attention obtained using attention distribution $a_i^t$. 
The \textit{copy} mechanism allows the model to locate a certain segment of the input sequence and puts that segment into the output sequence. 
$p_{cg}$ is a soft switch to choose between generating a word from vocabulary or copying a word from the input sequence.

\subsection{Incorporating a Coverage Mechanism}
The \textit{coverage} mechanism~\cite{tu2016modeling} is an improved method for the \textit{attention} mechanism. 
In the standard attention mechanism, the decoder dynamically assigns attention weights to different parts of the input sequence at each decoding step. 
However, as the decoding progresses, the \textit{attention} mechanism tends to focus on the same regions repeatedly, neglecting other parts of the input sequence. 
This can result in redundant or incomplete information being generated during the decoding process. 
To address this repetition problem, we incorporated the 
\textit{coverage} attention mechanism into our model.
Particularly, we maintain a word coverage vector ${cov}$ which is the sum of attention distributions over all previous decoder timesteps:
\begin{equation}
{cov}^{t}= \sum_{t'=0}^{t-1} a^{t'}.
\end{equation}
Here, ${cov}^t$ is a distribution over input tokens that represents the degree of coverage that those tokens have received from the \textit{attention} mechanism so far. 
Since no word is generated before timestamp 0, ${cov}^0$0 will be a zero vector. 
The update equation~\ref{equ:attention} is now modified to be:
\begin{equation}
    e^{t}_{i} = v^{t}\textnormal{tanh}\left(W_{cv}cov_{i}^{t} + W_{eh}h_{i} + W_{sh}s_{t} + b_{att} \right)
\end{equation}

Here, $W_{cv}$ are trainable parameters that ensure the \textit{attention} mechanism's current decision is informed by a reminder of its previous decisions.
The \textit{coverage} mechanism allows our model to solve the word repetition problem in the output sequence. 
The \textit{coverage} mechanism ensures that the \textit{attention} mechanism's current decision is informed by a reminder of its previous decisions (summarized in $cov^{t}$). 
This should make it easier for the \textit{attention} mechanism to avoid repeatedly attending to the same locations, and thus avoid generating repetitive text.
Following the incorporation of the \textit{copy} and \textit{coverage} mechanism in our sequence-to-sequence architecture, the final loss function will be:
\begin{equation}
    Loss = \frac{1}{T} \sum^{T}_{t=0}logP^{*}(Cword_{t}) + \lambda L_{cov}
\end{equation}
where $\lambda$ is a reweighted hyperparameter and the coverage loss $L_{cov}$ is defined as:
\begin{equation}
    L_{cov} = \sum_{i}min(a_{i}^{t}, cov_{i}^{t})
\end{equation}
Once the model is trained, we do inference using a beam search. 
The beam search is parametrized by the possible paths number $k$. 
The inference process stops when the model generates the \texttt{END} token which stands for the end of the sentence.


\section{Evaluation}
\label{setup}

\subsection{DataSet Preparation}
In this research, we reuse the raw smart contract dataset provided by Chen et al.~\cite{chen2021smart}, which contains 54,739 verified smart contracts.  
We describe how we prepared the dataset for our bytecode comment generation as follows. 

\subsubsection{\textbf{Data Collection.}}
For each verified smart contract, we crawled its source code and EVM bytecode from Etherscan~\cite{Etherscan}. 
As a result, we obtained 54,739 $\langle$srccode, bytecode$\rangle$ pairs of verified smart contracts. 
\begin{table}
  \caption{The Statistics of Our Collected Smart Contracts}
  \label{tab:dataset}
  \begin{tabular}{ccccc}
    \toprule
    Contract&Function&Comment&Bytecode Length Avg&Coment Length Avg\\
    \midrule
    54,739&1,323,554&565,403&6920.21&142.00\\
    \bottomrule
\end{tabular}
\end{table}
\begin{figure}[]\vspace{-1mm}
  \centering
  \includegraphics[width=0.85\linewidth]{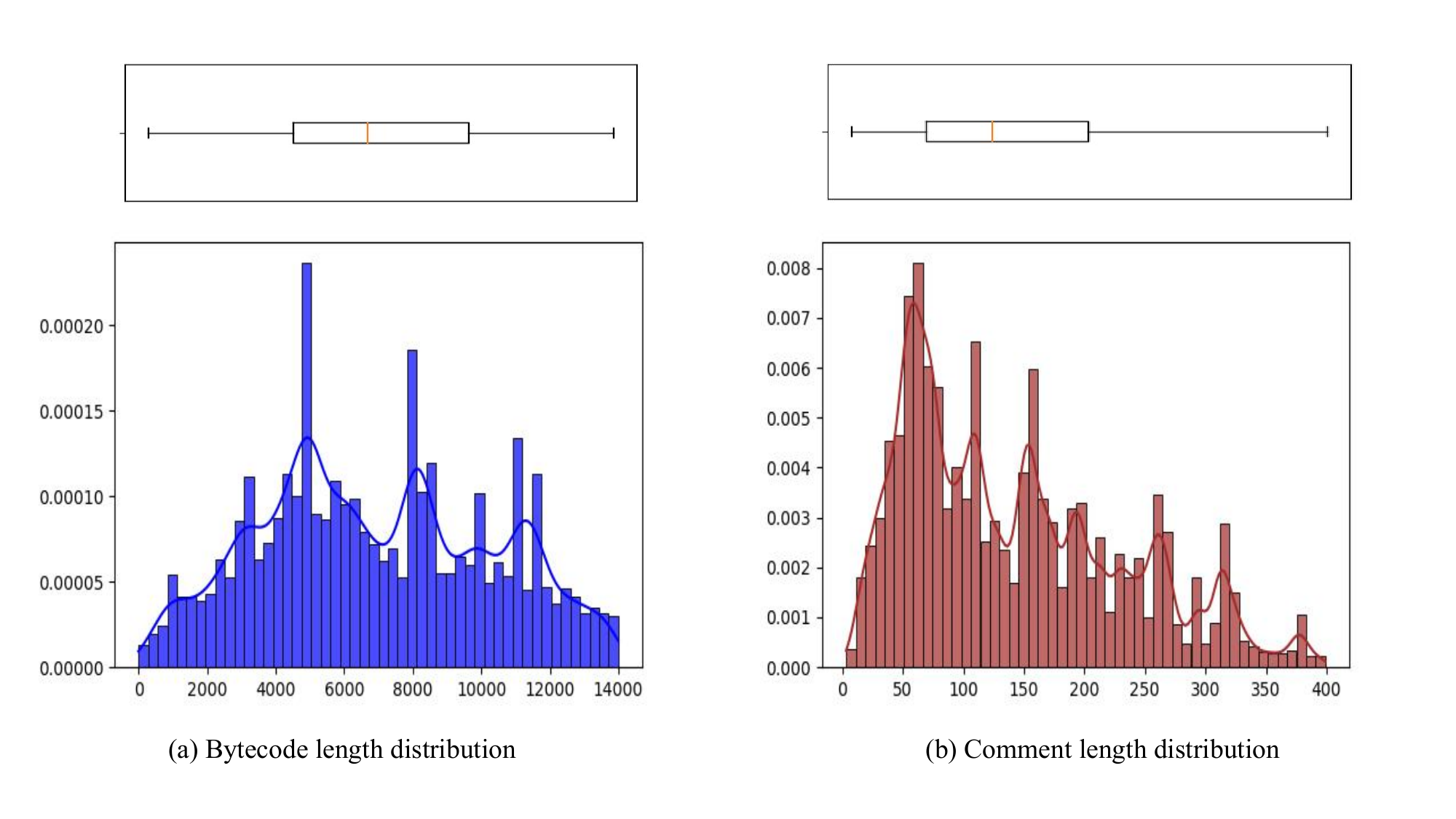}\vspace{-3mm}
  \caption{Length Distribution of The Training Data}\vspace{-3mm}
  \label{fig:dataset}
\end{figure}
\subsubsection{\textbf{Data Preprocessing.}}
We exploit the Solidity-parser to parse smart contract source code and extract functions within each collected smart contract. 
We define several regular expressions to extract smart contract comments for functions. 
Following that, we utilize the {\tt evm\_cfg\_builder} tool~\cite{EvmCfgBuilder} tool to generate CFG for each function from EVM bytecode. 
Our dataset now is constructed as $\langle$fucntion srccode,  function bytecode, function comment$\rangle$ triplets. 
Considering that duplicated data samples between the training set and testing set can mislead the evaluation results. 
We deduplicate our dataset according to the unique hash value generated by CFG and comment. 
Finally, we obtain 30,742 $\langle$function srccode, function bytecode, function comment$\rangle$ data samples. 
We list the statistics in Table~\ref{tab:dataset} and the length distribution of the bytecode and comment in Figure~\ref{fig:dataset}.

\subsubsection{\textbf{Data Splitting.}}
We split the constructed data samples into three chunks: 80 percent of the triple samples are used for training, 10 percent are used for validation and the rest are held out for testing. 
The training set is used to adjust the parameters, while the validation set is used to minimize overfitting, and the testing set is used only for testing the final solution to confirm the actual predictive power of our model with optimal parameters. 
The number of training, validation, and test sets of data samples are shown in Table~\ref{tab:split}.

\begin{table}
  \caption{Data Splitting Statistics}
  \label{tab:split}
  \begin{tabular}{cccl}
    \toprule
    Total&Tain&Validate&Test\\
    \midrule
    30,742&24,594&3074&3074\\
    \bottomrule
\end{tabular}
\end{table}

\subsection{Evaluation Metrics}
To demonstrate the effectiveness of \appname, we use two widely used metrics in comment and code generation tasks~\cite{iyer2016summarizing}: 
\subsubsection{\textbf{BLEU}}
BLEU~\cite{papineni2002bleu} is a precision-oriented measure, which measures the average $n$-gram precision on a set of reference sentences, with a penalty for overly short sentences. 
It is widely used in various tasks of automatic software engineering, such as API sequence generation~\cite{gu2016deep}, comment generation~\cite{hu2018deep, hu2020deep, wei2020retrieve}, and commit message generation~\cite{jiang2017automatically}. 
BLEU calculates the similarity between the generated notice and references.
The similarity is computed as the geometric mean of n-gram matching precision scores multiplied by a brevity penalty to prevent very short generated sentences. 
In addition, we introduce the smoothing function Smooth2 in BLEU evaluation, which can improve the stability of BLEU scores and reflect translation quality more accurately. 
In this paper, we adopt BLEU-1, BLEU-2, BLEU-3, and BLEU-4 scores. 


\subsubsection{\textbf{ROUGE}}
ROUGE~\cite{lin2004rouge} is a widely used recall-oriented measure in summarization tasks. 
It evaluates the overlap of $n$-grams between system-generated summaries and reference sentences. 
ROUGE-1 and ROUGE-2 measure unigram and bigram overlaps, while ROUGE-L captures in-sequence matches reflecting sentence-level word order. To evaluate different models, we consider ROUGE-1, ROUGE-2, and ROUGE-L scores.

\subsection{Training Details}
We implement \appname on top of Pytorch.
Both token embeddings and hidden size are set to 256 dimensions. 
All parameters are optimized using Adam~\cite{kingma2014adam} with the initial learning rate of 0.0005.
Following Vaswani et al.~\cite{vaswani2017attention}, we increase the learning rate linearly for the first 4000 steps (i.e., warmup steps) and decrease it thereafter proportionally to the inverse square root of the step number.
During the training, the batch size is set to 32.
To mitigate overfitting, we exploit the dropout mechanism and set the dropout rate as 0.1.
We set the maximum length of the encoder to 200 and the maximum length of the decoder to 50. Training runs for 50 epochs.
We conduct our experiments on a Linux server with an NVIDIA GeForce RTX 2080Ti GPU having 10 GB memory.

\section{Experiment Results}
\label{results}

In this study, we aim to answer the following research questions:
\begin{itemize} 
    \item \textit{RQ1:} How effective is our \appname for generating smart contract comments from bytecode?
    \item \textit{RQ2:} How effective is the IR component with different retrieval methods?
    \item \textit{RQ3:} How effective is our use of \textit{attention} mechanism, \textit{copy} mechanism and \textit{coverage} mechanism under automatic evaluation? 
    \item \textit{RQ4:} How effective is our \appname under different IR settings?
    \item \textit{RQ5:} How effective are baseline models augmented with IR component?
    \item \textit{RQ6:} How effective are LLMs for generating smart contract comments from bytecode?
    \item \textit{RQ7:} How effective is our \appname under human evaluation?
\end{itemize}

\subsection{RQ1. \appnamebold Overall Effectiveness}
\subsubsection{Experimental Setup.}
In this RQ, we want to investigate how effective our approach and baseline methods are for generating smart contract comments from bytecode. 
In particular, all models are trained (i.e., \appname) and fine-tuned (i.e., CodeT5 and PLBART) with our training set. 
Then the models (including \appname and baselines) with their best performance on the validation set are used to report final comparison results.  
For each smart contract function in our test set, we use the intermediate representation of bytecode, i.e., CFG sequence, as model inputs (including \appname and baseline methods). 
Then \appname and baseline methods generate corresponding function comments as outputs based on the inputs. 
We then calculate the BLEU and ROUGE scores between the generated comments and ground truth comments for comparison purposes.

\subsubsection{Baselines.}
Since no prior work focused on contract comment generation based on bytecode, we chose several commonly used baselines for this task. 
The following baselines are adopted in this study: 

\begin{itemize} 
    \item \textbf{IR}: 
    The IR stands for information retrieval baseline. 
    For a given bytecode, it retrieves a function comment that is closest to the input bytecode from the training set. 
    Since directly comparing bytecode is difficult, we 
    calculate the similarity between the CFG sequence of two different bytecodes, and the most similar comment is fetched as the IR method result. 
    In this study, we employ widely used  BM25~\cite{robertson2009probabilistic} as our information retrieval algorithm to perform the information retrieval task. 
    
    \item \textbf{CodeT5}: 
    CodeT5~\cite{wang2021codet5} is a unified pre-trained encoder-decoder Transformer model.  
    CodeT5 better leverages the code semantics conveyed from the developer-assigned identifiers. 
    CodeT5 builds on the T5 architecture~\cite{raffel2020exploring} that supports both code understanding and generation and allows for multi-task learning. 
    CodeT5 outperforms prior methods in code generation tasks including PL-NL, NL-PL and PL-PL. 
    In this work, we use CodeT5 (CodeT5-small and CodeT5-base) as a baseline to perform our comment generation tasks. 
    Considering it is difficult for CodeT5 to directly learn from bytecode, we use CFG sequences as inputs for CodeT5. 
    We then fine-tuned the pre-trained CodeT5 model with our training set on the code summarization task, using CFG sequences as inputs and target corresponding comments as outputs. 
    We fine-tuned CodeT5 for 50 epochs and the model with the highest BLEU-4 score was chosen for evaluation. 
    
    \item \textbf{PLBART}: 
    PLBART~\cite{ahmad2021unified}
    is a bidirectional and autoregressive transformer pre-trained on unlabeled data across PL (Programming Language) and NL (Natural Language) to learn multilingual representations. 
    PLBART is pre-trained on an extensive collection of Java and Python functions and NL text via denoising autoencoding. 
    PLBART outperforms or rivals state-of-the-art models on code summarization, code generation, and code translation tasks. 
    Similarly, we use CFG sequences as PLBART inputs. 
    We fine-tuned PLBART the same way of fine-tuning CodeT5, the best model on the validation set is used for final evaluation. 

\end{itemize}

\subsubsection{Evaluation Results.}
The evaluation results of \appname and aforementioned baselines are summarized in Table~\ref{tab:evaluation}. 
From the table, we can observe the following points:
\begin{enumerate}
    \item \textbf{In general, the encoder-decoder architecture baselines (i.e., CodeT5 and PLBART) outperform IR based approach (i.e. BM25)}. 
    For IR based approach, it retrieves the comments from the existing database according to similarity scores, which heavily relies on the presence of similar Control Flow Graphs (CFGs) and the degree of similarity between them, indicating that merely memorizing the training set is not enough for this task. 
    In contrast to the IR-based approach, the encoder-decoder model uses the vector representation for tokens and internal states, semantic and structural information can be learned from these vectors by taking global context into consideration. 
    \item \textbf{Notably, our IR-augmented model (i.e., {\sc SmartBT}) even achieves a better performance than the pre-trained transformer-based models (e.g., CodeT5 and PLBART)}. 
    This is because these pre-trained language models are mostly pre-trained on the extensive collection of code corpus (e.g., source code and NL text), however, these pre-trained models were not designed to handle opcode (e.g., CFG sequence), which results in their suboptimal performance for our newly proposed task. 
    Moreover, compared with source code, the gap between the bytecode/opcode and comments is even larger, posing a significant challenge for pre-trained models to capture their semantic relationships effectively. 
    This is the reason why we introduce IR-augmented techniques for bridging the semantic gap between low-level bytecode and natural language comments. 
    \item \textbf{Regarding BLEU score, our approach is significantly better than other baselines and achieves understandable results.} 
    For example, it improves over PLBART on BLEU-4 by 22\%. 
    We attribute this to the following reasons: besides solely depending on structural inputs such as CFGs, our model also incorporates an IR-augmented module, the IR augmented module fetches comments from similar smart contracts. 
    In other words, {\sc SmartBT} combines the structural input (i.e., CFGs) and semantic input (i.e., similar comments) for constructing the contextual vectors, which signals that similar comments convey much valuable information when generating target comments. 
    \item \textbf{Regarding ROUGE score, the advantage of our approach is also clear.} This is because our model is enhanced by a \textit{copy} mechanism to handle rare-word problems as well as a \textit{coverage} mechanism to eliminate meaningless repetitions. 
    This further justifies the aforementioned mechanisms generally help when dealing with the comment generation task. 
\end{enumerate}

\find{
\textbf{Answer to RQ-1: How effective is our \appname for generating smart contract comments from bytecode?} 
We conclude that our approach is effective for generating smart contract comments from the bytecode under automatic evaluation and surpasses baselines by a large margin. 
}

\begin{table}[]\vspace{-3mm}
  \caption{Automatic Comment Generation Evaluation Using BLEU and ROUGE Scores (\%) }\vspace{-3mm}
  \label{tab:evaluation}
   \resizebox{\linewidth}{!}{
    \begin{tabular}{lrrrrrrr}
    \toprule
    Method&BLEU-1&BLEU-2&BLEU-3&BLEU-4&ROUGE-1&ROUGE-2&ROUGE-L\\
    \midrule
    BM25&24.08&23.13&22.57&22.24&26.35&20.32&25.78\\
    CodeT5-small&25.49&24.69&24.11&23.55&29.49&21.80&29.02\\
    CodeT5-base&28.18&27.44&26.83&26.21&32.10&24.76&31.67\\
    PLBART-base&30.03&29.01&27.93&27.24&36.01&27.80&35.46\\
    \midrule    
    
    \appnamebold &\textbf{37.18}&\textbf{35.65}&\textbf{34.15}&\textbf{33.24}&\textbf{41.14}&\textbf{32.44}&\textbf{40.53}\\
    \bottomrule
  \end{tabular}
  }
\end{table}


\subsection{RQ2. Effectiveness of Different IR Methods}
\subsubsection{Experimental Setup.} 
In this RQ, we want to investigate how much performance improvement our approach can achieve by using the IR-augmented component. 
In particular, we selected several different methods as the information retrieval algorithm for IR-augmented component. 
We then used these different IR-augmented components to generate target comments and calculated the BLEU and ROUGE scores for comparison purposes. 
\subsubsection{Baselines.}
To fill the semantic gap between bytecode and comment, we introduce the IR (Information-retrieval) augmented module to fetch similar comments from contracts with similar code structures and logic. 
Intuitively, our IR augmented module can take any similarity matching methods. 
To verify the effectiveness of our using IR augmented component, we choose the following information retrieval methods for this research question. 
\begin{itemize}

    \item $\mathbf{IR_{None}}$: For this baseline, we remove the IR augmented component and keep the rest of {\sc SmartBT}. 
    In other words, we drop the semantic input (i.e., similar comment) of our model and only retain the structural input (i.e. CFGs). 
    This baseline is denoted as $\mathbf{IR_{None}}$. 

    \item \textbf{BOW:} 
    Cosine similarity~\cite{salton1988term} is one of the most popular distance metrics used for comparing the similarity between two vectors. 
    Regarding this baseline, we convert the CFG sequence into BOW (bag-of-words) vectors, then cosine similarity scores are computed between any two given CFGs, this baseline is denoted as \textbf{BOW}. 



    \item \textbf{GraphCodeBERT:} 
    Besides using the BOW to represent the CFG sequence, we also use GraphCodeBERT~\cite{guo2020graphcodebert} to encode CFG into vector representations. 
    In particular, for a given smart contract bytecode, its CFG sequence is fed into GraphCodeBERT to obtain the feature representation vector. 
    Then the vector is used to fetch the most similar smart contract by calculating the cosine similarity scores between two embedding vectors. 
    \item {\sc \textbf{SmartBT:}} 
    Our approach employs the widely used BM25 as our information retrieval algorithm. 
    BM25 aims to provide accurate and relevant search results by scoring documents based on their term frequencies and document lengths.
    It follows the probabilistic retrieval framework, which assumes that relevant and non-relevant documents follow different statistical distributions.
\end{itemize}
\begin{table}[h]
    \caption{Effects of Different Information Retrieval Methods on Our Task: BLEU and ROUGE Scores (\%)}
    \vspace{-3mm}
    \label{tab:information_retrieval}
    \resizebox{\linewidth}{!}{
    \begin{tabular}{lccccccc}
    \toprule
    Method&BLEU-1&BLEU-2&BLEU-3&BLEU-4&ROUGE-1&ROUGE-2&ROUGE-L\\
    \midrule
    $\mathbf{IR_{None}}$&31.18&30.19&29.36&28.90&34.61&27.46&34.03\\
    \textbf{Bag-Of-Words}&33.22&32.51&31.11&30.00&37.21&30.47&36.89\\
    \textbf{GraphCodeBERT}&33.66&32.96&31.65&30.66&39.18&32.80&38.85\\
    \midrule
    \appnamebold &\textbf{37.18}&\textbf{35.65}&\textbf{34.15}&\textbf{33.24}&\textbf{41.14}&\textbf{32.44}&\textbf{40.53}\\
    \bottomrule
  \end{tabular}
  }
\end{table}

\subsubsection{Evaluation Results.}The evaluation results of \appname and aforementioned baselines are summarized in Table~\ref{tab:information_retrieval}. From the table, it can be seen:
\begin{itemize} 
    \item[(1)] 
    \textbf{The IR augmented module is effective in enhancing the effectiveness of our model and contributes to the overall performance.} Compared with $\mathbf{IR_{None}}$, there is an improvement in terms of all evaluation metrics after adding IR augmented module, regardless of the information retrieval algorithm. 
    For example, using \textbf{BOW} and \textbf{GraphCodeBERT} to fetch relevant comments as inputs, the BLEU-4 score has improved by 3.8\% and 6\% respectively. 
    The experimental results indicate that the encoder-decoder neural network merely using structural inputs (i.e., CFGs) is unable to bridge the semantic gap between bytecode and comments, verifying the importance and necessity of using semantic inputs (i.e., similar comments) via using IR augmented component.  
    \item[(2)] \textbf{\appnamebold, using the BM25 algorithm, achieved the best performance among other information retrieval algorithms.} We attribute this to the following advantages of the BM25 algorithm: (a) Term saturation: BM25 incorporates a term saturation function, this function mitigates the impact of excessively high term frequencies. (b) Dynamic ranking: BM25 adjusts its ranking based on the distribution of terms within the collection, making it more adaptable to different types of documents and queries. 
    (c) Effective for long queries: BM25 is effective in handling long CFG sequences as it addresses the issue of term saturation and considers the overall CFG sequence length. 
\end{itemize}

\find{
\textbf{Answer to RQ-2: How effective is the IR component with different methods?} 
We conclude that the inclusion of the IR augmented module significantly enhances our model's effectiveness and contributes to its overall performance. 
The BM25 performs best for fetching relevant comments from smart contracts and provides semantic inputs for \appnamebold.  
}


\subsection{RQ3. Effectiveness of Using \textit{Copy} and \textit{Coverage} Mechanisms}
\subsubsection{Experimental Setup.}
In this RQ, we conduct an ablation experiment to verify the effectiveness of using the \textit{copy} mechanism and \textit{coverage} mechanism within our model. 
In particular, we removed the \textit{copy} mechanism and \textit{coverage} mechanism one by one, and then we calculated the BLEU and ROUGE scores between the generated comments and ground truth comments for comparison purposes.

\subsubsection{Baselines.} 
When constructing \appnamebold, we added an \textit{attention} mechanism, a \textit{copy} mechanism, and a \textit{coverage} mechanism to our IR augmented encoder-decoder architecture. 
The \textit{attention} mechanism is often regarded as a basic mechanism incorporated within the sequence-to-sequence model. 
The \textit{copy} mechanism is used to handle rare-word problems in our generation process and the \textit{coverage} mechanism is used to eliminate meaningless repetitions while decoding. 
To verify the effectiveness of the above mechanisms, to be more specific, we compare our \appnamebold with several of its incomplete variants: 
\begin{itemize}
    \item $\textbf{Model}_{atten}$: This baseline removes the \textit{copy} and \textit{coverage} mechanisms from our approach while keeping the basic \textit{attention} mechanism as the basic model. 
    \item $\textbf{Model}_{atten+copy}$: This baseline removes the \textit{coverage} mechanism from our approach while keeping the \textit{attention} mechanism and the \textit{copy} mechanism. 
    \item $\textbf{Model}_{atten+coverage}$: This baseline removes the \textit{copy} mechanism from our approach while keep the \textit{attention} mechanism and \textit{coverage} mechanism.
    \item \appnamebold: It is our current work which considers all the three mechanisms. 
\end{itemize}

\begin{table}[h] \vspace{-2mm}
  \caption{Ablation Evaluation of Different Mechanisms Using BLEU and ROUGE Scores (\%)}\vspace{-3mm}
  \label{tab:ablation}
   \resizebox{\linewidth}{!}{
   \begin{tabular}{lccccccc}
    \toprule
    Model&BLEU-1&BLEU-2&BLEU-3&BLEU-4&ROUGE-1&ROUGE-2&ROUGE-L\\
    \midrule
    $\textbf{Model}_{atten}$&28.37&27.23&26.32&25.83&32.01&24.29&31.39\\
    $\textbf{Model}_{atten+copy}$&32.29&31.18&30.28&29.81&35.95&28.32&35.38\\
    $\textbf{Model}_{atten+coverage}$&30.11&28.84&27.90&27.36&33.32&25.59&32.70\\
    \midrule
    \appnamebold&\textbf{37.18}&\textbf{35.65}&\textbf{34.15}&\textbf{33.24}&\textbf{41.14}&\textbf{32.44}&\textbf{40.53}\\
    \bottomrule
  \end{tabular}
  }
\end{table}
\subsubsection{Evaluation Results.}
\begin{figure}[h]
  \centering
  \includegraphics[width=0.95\linewidth]{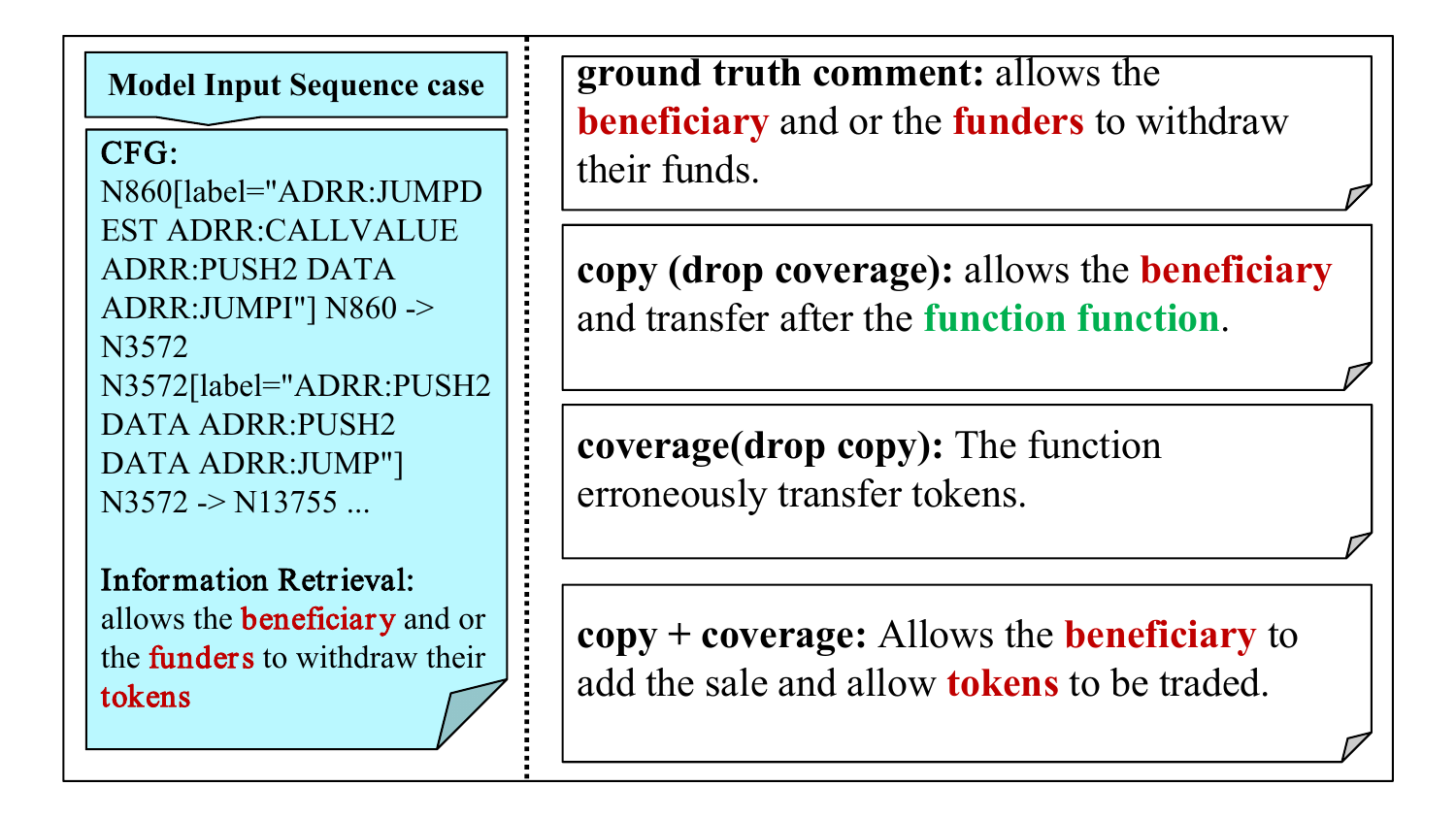}\vspace{-3mm}
  \caption{Ablation Analysis Example}\vspace{-3mm}
  \label{fig:ablation}
\end{figure}

The evaluation results of the ablation study are summarized in Table~\ref{tab:ablation}. 
From the table, we can observe the following points:
\begin{itemize} 
    \item[(1)] By comparing the results of $\textbf{Model}_{atten}$ with $\textbf{Model}_{atten+copy}$ and $\textbf{Model}_{atten+coverage}$, we can measure the performance improvements achieved due to the incorporation of \textit{copy} mechanism and \textit{coverage} mechanism respectively. 
    \textbf{It is clear that better performance can be achieved by solely adding \textit{copy} or \textit{coverage} mechanism to the attention based model}. 
    This signals that both \textit{copy} and \textit{coverage} mechanism do have contributions to the overall performance improvements. 
    \item[(2)] By comparing the results of our approach and each of the variant model, \textbf{we can see that no matter which type of mechanism we removed, there is a drop overall in every evaluation measure metric and does hurt the performance of our model.} 
    Particularly, when comparing the our model \appnamebold with $\textbf{Model}_{atten}$, it drops almost 10\% of the overall automatic evaluation scores. This verifies the importance and effectiveness of these incorporated mechanisms. 
    \item[(3)] To gain insights into these mechanisms, we further illustrate an example in Figure~\ref{fig:ablation} from the ablation analysis to show the effect of employing \textit{copy} and \textit{coverage} mechanism. 
    From the figure, we can see that: (a) After introducing the \textit{copy} mechanism, the model can \textbf{copy} relevant tokens from our IR augmented module to the generated comment. For example, the words \texttt{beneficiary} and \texttt{funders} (colored in red) are copied directly from the IR-augmented component to the target outputs. (b) Repetition is a common problem for attentional sequence-to-sequence models, meaningless repeated words are produced during the generation process (highlighted with green color). The \textit{coverage} mechanism is effective for discouraging such repetitions by quantitatively emphasizing the coverage of sentence words while decoding. For example, the word ``\texttt{function}'' has been meaningless repeated twice, employing \textit{coverage} mechanism can effectively eliminate such repetitions.  
\end{itemize}

\find{
\textbf{Answer to RQ-3: How effective is our use of \textit{attention} mechanism, \textit{copy} mechanism and \textit{coverage} mechanism under automatic evaluation?} We conclude that all the incorporated mechanisms do have contributions to the overall performance and are effective and helpful in enhancing the performance of \appnamebold. 
}

\subsection{RQ4. Effectiveness of Different IR Settings}
\subsubsection{Experimental Setup.}
In this RQ, we want to investigate how \appnamebold performs under different IR augmented module settings. 
In particular, we aim to explore the optimal IR-augmented module settings for our task. 
Based on the findings of RQ3, we vary the number of retrieved comments for our model inputs and compare generated results under different module settings. 

\subsubsection{Baselines.} 
The key hyperparameter of our IR augmented module is the number of similar comments retrieved for \appnamebold inputs. 
In this RQ, to investigate the optimal settings for the IR augmented module, we vary the number of retrieved comments to construct the baselines, denoted as $\textbf{IR}_k$, where $k$ represents the number of retrieved comments. 
In other words, $\textbf{IR}_k$ retrieves $k$ most relevant comments from our database and concatenates these comments as semantic inputs for our model. 
The $k$ is varied in $[0, 1, 3, 5]$ for this study, where $k=1$ represents our current model \appnamebold. 

\begin{table}[h]\vspace{-3mm}
  \caption{Effects of Different Settings of IR Component Using BLEU and ROUGE Scores (\%)}\vspace{-3mm}
  \label{tab:ir_k}
   \resizebox{\linewidth}{!}{
   \begin{tabular}{lccccccc}
    \toprule
    Model&BLEU-1&BLEU-2&BLEU-3&BLEU-4&ROUGE-1&ROUGE-2&ROUGE-L\\
    \midrule
    $\textbf{IR}_{k=0}$&31.18&30.19&29.36&28.90&34.61&27.46&34.03\\
    $\textbf{IR}_{k=3}$&26.23&25.33&24.16&23.22&30.54&22.48&30.01\\
    $\textbf{IR}_{k=5}$&23.78&22.13&21.03&20.34&27.60&17.41&26.91\\
    \midrule
    \appnamebold&\textbf{37.18}&\textbf{35.65}&\textbf{34.15}&\textbf{33.24}&\textbf{41.14}&\textbf{32.44}&\textbf{40.53}\\
    \bottomrule
  \end{tabular}
  }
\end{table}

\subsubsection{Evaluation Results.}
The evaluation results are listed in Table~\ref{tab:ir_k}, it can be seen that: 
\begin{itemize}
    \item[(1)] \textbf{Adding semantic input for our model greatly improves the overall performance of our model.} 
    Comparing our approach to the case of $\textbf{IR}_{k=0}$ (where no similar comments are retrieved for semantic inputs), our model shows an improvement of nearly 19\% in overall evaluation metrics. 
    This demonstrates the significant impact of our IR-augmented module and the appropriate value of $k$. 
    \item[(2)] \textbf{Regarding the number of retrieved comments, not the more the better.}
    When adding more retrieved comments to our semantic inputs (e.g., top3 and top5 similar comments), we can find that the BLEU and ROUGE scores significantly drop. 
    This is because a larger $k$ may introduce more noise into our semantic inputs with more irrelevant references, which can increase the difficulty of generating correct comments. 
    \item[(3)] \textbf{By analyzing the performance of our approach with respect to different $k$, we notice that \appnamebold achieves its optimal performance when $k=1$.} The overall performance trend of our model decreases as $k$ increases, which supports our concern that larger $k$ settings introduce more noise and bring bigger challenges for our task. 
    We thus set $k=1$ for constructing our IR augmented module. 
\end{itemize}


\find{
\textbf{Answer to RQ-4: How effective is our \appnamebold under different settings of IR component?} We conclude that our \appnamebold performs best when the top 1 relevant comment is retrieved for our semantic input, further increasing the $k$ can introduce more noise and bring bigger challenges for our task. 
}

\subsection{RQ5. Effectiveness of Adding IR Component on Baselines}
\subsubsection{Experimental Setup.} 
In this RQ, we aim to explore the impact of IR component on enhancing the performance of baseline models. 
Specifically,  we add the same IR component incorporated within our model to baselines. 
For a fair comparison, the information retrieval methods (i.e., BM25) and its settings (i.e., number of retrieving comments) are exactly the same with \appname. 
Following that, we augmented baselines with IR component to generate target comments and calculate the BLEU and ROUGLE scores for evaluation. 

\subsubsection{Baselines.} 
In RQ1, we adopted the Pre-trained Language Models, i.e., CodeT5 (small and base models) and PLBART, as baselines for comparison. 
We thus augment the above pre-trained models with the IR component, denoted as \textit{CodeT5-small+IR}, \textit{CodeT5-base+IR} and \textit{PLBART-base+IR} respectively. 
We provide the performance of each model without an IR component for easy reference. 
In this research question, we also provide the parameter size as an indicator for reference.

\subsubsection{Evaluation Results.}
The experimental results are shown in Table~\ref{tab:irbaselines}, from the table we can observe the following points:
\begin{enumerate}
    \item By comparing the models augmented with IR components and without IR components, \textbf{it is clear that IR augmented models exhibit significant improvements over their respective baseline models across all evaluation metrics, suggesting considerable advancements for introducing the IR component}. 
    For example, after adding the IR component to \textit{CodeT5-small} and \textit{CodeT5-base} models, their BLEU scores have been improved by 43\% and 48\% respectively. 
    The ROUGE scores also demonstrate substantial improvements, \textit{CodeT5-small+IR} and \textit{CodeT5-base+IR} models show a notable improvement of 38\% and 43\% respectively, while \textit{PLBART-base+IR} model also achieves an improvement of 7\%. 
    Overall, the introduction of the IR component can greatly enhance the performance of baseline models, further verifying the effectiveness of the IR component in filling the semantic gap and improving the quality of generated comments. 
    \item After adding the IR component, the \textit{CodeT5-base+IR} model even outperforms our \textsc{SmartBT} in terms of BLEU and ROUGE scores, while \textbf{our model has its advantages regarding parameter sizes, which is much more lightweight and flexible}. 
    Compared with our \textsc{SmartBT}, our model only costs 26M, while \textit{CodeT5-base+IR} and \textit{PLBART-base+IR} cost 220M and 1,100M respectively, which are 
    8 and 40 times larger than our model. 
    Our \textsc{SmartBT} can be easily set up and deployed on developers' personal computers, it can perform inferencing with CPUs, without requiring GPUs. 
    As a result, our proposed model is more lightweight and flexible according to different developers' computing resources. 
\end{enumerate}

\begin{table}[]\vspace{-3mm}
  \caption{Effects of IR Augmented Baselines Using BLEU and ROUGE Scores (\%)}\vspace{-3mm}
  \label{tab:irbaselines}
   \resizebox{\linewidth}{!}{
    \begin{tabular}{lrrrrrrrr}
    \toprule
    Method&BLEU-1&BLEU-2&BLEU-3&BLEU-4&ROUGE-1&ROUGE-2&ROUGE-L&\# Parameters\\
    \midrule
    CodeT5-small&25.49&24.69&24.11&23.55&29.49&21.80&29.02&60M\\
    CodeT5-small+IR&36.53&34.29&33.01&32.22&40.71&32.90&40.17&60M\\
    CodeT5-base&28.18&27.44&26.83&26.21&32.10&24.76&31.67&220M\\
    CodeT5-base+IR&41.93&39.96&38.54&37.66&45.82&38.88&45.40&220M\\
    PLBART-base&30.03&29.01&27.93&27.24&36.01&27.80&35.46&1100M\\
    PLBART-base+IR&35.39&29.30&26.80&25.80&38.14&29.03&36.70&1100M\\
    \midrule    
    \appnamebold &\textbf{37.18}&\textbf{35.65}&\textbf{34.15}&\textbf{33.24}&\textbf{41.14}&\textbf{32.44}&\textbf{40.53}&\textbf{26M}\\
    \bottomrule
  \end{tabular}
  }
\end{table}


\find{
\textbf{Answer to RQ-5: How effective are baseline models augmented with IR component?} 
The IR-augmented models show significant improvements over baseline models across all evaluation metrics, further verifying the effectiveness and importance of introducing IR component.} 

\subsection{RQ6. The Effectiveness of LLMs on Our Tasks}


\subsubsection{Experimental Setup.}
The Large Language Models (LLMs) (e.g., ChatGPT) are widely used by developers nowadays and have demonstrated promising performance for code-related tasks, such as code generation, test generation, and code summarization~\cite{mai2024human, dai2024mpcoder, yan2023closer, xue2024selfpico, wang2024makes}. 
In this RQ, we want to investigate whether the LLMs can be adapted to our task successfully. 
In particular, we aim to explore the ability of LLMs on our tasks in terms of two aspects: generating comments directly from smart contract bytecode and generating comments from CFG sequences. 
Since it is too expensive to conduct the experiments on our full test set (i.e., 3,074 cases), we randomly selected a statistically representative sample of 342 cases from our test set (with a 95\% confidence level and 5\% margin of error). 
Following that, we calculated the BLEU and ROUGE scores of LLMs and \textsc{SmartBT} and these 342 cases for comparison purposes.

\subsubsection{Baselines}
\begin{figure}[t]
  \centering
  \includegraphics[width=0.9\linewidth]{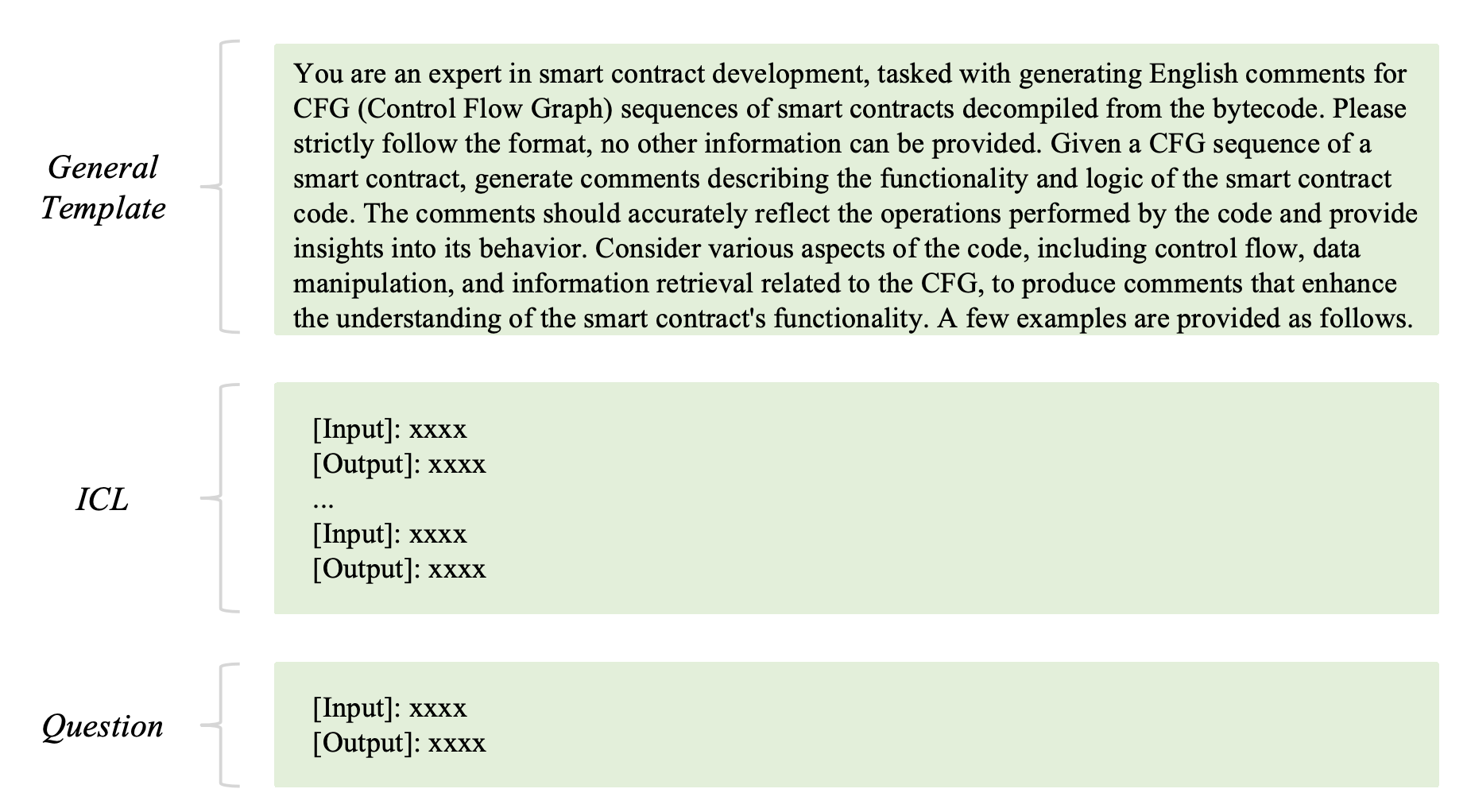}
  \caption{An ICL Prompt Template Example}
  \label{fig:iclcase}
\end{figure}

The great success of ChatGPT demonstrates the remarkable ability of large
language models (LLMs) to comprehend human questions and assist in code-relevant tasks. 
In this RQ, we adopted the GPT-4 as the baseline to perform our task. 
GPT-4 is the newest LLM created by OpenAI, it is a large multimodal model (accepting image and text inputs and emitting text outputs), that exhibits human-level performance on various professional and academic benchmarks. 
Currently, in-context learning (i.e., ICL) has been widely used by researchers to elicit human knowledge and logical reasoning from LLMs to accomplish complicated tasks. 
We explore the performance of two variants of in-context learning (i.e., zero-shot learning and few-shot learning) on two scenarios (i.e., input Bytecode and CFG sequence) respectively. 
We set the temperature request parameter to 0.7, which is a commonly used empirical standard in many papers~\cite{guo2023large}. 
This setting is known to generate results that are relatively natural and exhibit a certain level of creativity. 
\begin{itemize}
    \item \textbf{Zero-shot Learning:} With the increasing ability of LLMs, in-context learning has shifted to a new paradigm, known as zero-shot learning, where LLMs make predictions by directly describing the desired output. 
    Zero-shot prompting involves posing a question or task to the model without providing any specific context or examples. 
    We apply zero-shot learning to bytecode and CFGs respectively, denoted as GPT-4-Zero-RBC and GPT-4-Zero-CFG.
    \item \textbf{Few-shot Learning:} While zero-shot learning shows promising performance in various tasks by leveraging prior knowledge from training sources, it remains challenging to apply to unseen tasks. 
    To overcome this challenge, few-shot learning is utilized to augment the context with a few examples of desired inputs and outputs (as shown in Fig.~\ref{fig:iclcase}). 
    Few-shot learning enables LLMs to recognize the input prompt syntax and patterns of output. 
    We apply few-shot learning to bytecode and CFG input respectively, denoted as  GPT-4-Few-RBC and GPT-4-Few-CFG. 
\end{itemize}

\begin{table}[ht]
  \caption{Effects of GPT-4 on Our Task Using BLEU and ROUGE Scores (\%)}
  \label{tab:llm}
    \resizebox{\linewidth}{!}{
        \begin{tabular}{lccccccccc}
        \toprule
        Method&BLEU-1&BLEU-2&BLEU-3&BLEU-4&ROUGE-1&ROUGE-2&ROUGE-L&\# Parameters\\
        \midrule
        GPT-4-Zero-RBC&10.56&8.00&7.00&6.41&14.24&4.33&13.63&1760B\\
        GPT-4-Few-RBC&13.64&11.02&9.84&9.04&18.42&8.05&17.48&1760B\\
        GPT-4-Zero-CFG&14.75&12.14&10.57&9.56&21.73&9.71&20.29&1760B\\
        GPT-4-Few-CFG&17.01&14.19&12.70&11.74&24.18&11.32&22.74&1760B\\
        \midrule
        \appnamebold &\textbf{37.18}&\textbf{35.65}&\textbf{34.15}&\textbf{33.24}&\textbf{41.14}&\textbf{32.44}&\textbf{40.53}&\textbf{26M}\\
        \bottomrule
        \end{tabular}
    }
\end{table}
\subsubsection{Evaluation Results.}
The experimental results of GPT-4 and \textsc{SmartBT} on the 342 test cases are summarized in Table~\ref{tab:llm}, from the table, several points stand out:
\begin{enumerate}
    \item \textbf{It is difficult for LLM to generate comments from bytecode directly.} 
    Compared with generating comments from CFG sequences, generating comments directly from bytecode inputs achieves the worst performance in terms of all evaluation metrics. 
    Our experimental results reveal that LLMs can hardly understand the semantic meaning embedded within the bytecode, this is reasonable because even advanced LLMs such as GPT-4, have not been pre-trained with bytecode data, thereby hindering them from generating target comments accurately and effectively. 
   
    \item By comparing GPT-4's performance with zero-shot learning and few-shot learning, it is clear that \textbf{few-shot learning has its advantage over zero-shot learning.} For example, with few-shot learning, GPT-4 improves the BLEU-4 score over zero-shot learning by 41\% and 22\% regarding bytecode and CFG sequence respectively. 
    This is because zero-shot learning relies solely on the model’s preexisting knowledge to generate responses. 
    When GPT-4 doesn't have specific knowledge triggered by the prompt, it may provide generic or unrelated responses. 
    Few-shot learning prompts the GPT-4 with concrete examples, enhancing the model's understanding of the given task. 

    \item  \textbf{\appname outperforms GPT-4 by a large margin in terms of all evaluation metrics}. 
    The suboptimal performance of GPT-4 on our tasks indicates that LLMs are not suitable for this task because they are not pre-trained with the bytecode datasets and/or designed for handling bytecode-relevant tasks. 
    Compared with GPT-4, we introduce CFGs to capture the structural information of bytecode and IR-augmented components to capture the semantic information of bytecode. 
    It would be interesting to explore the effectiveness of pre-training and fine-tuning LLMs with bytecode and/or developing LLMs particularly tailored to handle bytecode-related tasks, but this is beyond the scope of our current research and we plan to leave it for future work. 
\end{enumerate}

\find{
\textbf{Answer to RQ-6: How effective are LLMs for generating smart contract comments from bytecode?} 
We conclude that it is difficult for LLMs such as GPT-4 to generate comments from smart contract bytecode because LLMs are seldom pre-trained with bytecode data and can hardly generalize to this unseen task.
}

\subsection{RQ7. Human Evaluation}

\subsubsection{Experimental Setup.} 
Since automated evaluation indicators such as BLEU and ROUGE cannot really reflect the effects of generating comments, we conducted a user study to make more real evaluations. 
In this RQ, we investigate the effectiveness of our approach through human evaluation~\cite{gao2020generating}.  
To be more specific, we invited participants to manually assess the quality of our generated comments in terms of the following two aspects: 
\begin{itemize}
    \item \textbf{Naturalness:} Naturalness evaluates the grammatical correctness and fluency of the generated comments. It assesses how well the comments read and flow in natural language, ensuring they are easily comprehensible to human users.
    \item \textbf{Relevance:} Relevance assesses the comments' alignment with the ground truth comment. It measures the degree of relevance and coherence between the generated and reference comments, indicating how well the generated comments capture the intended meaning and information present in the original code. 
\end{itemize}

\subsubsection{Baselines.} 
\begin{figure}[t]\vspace{-3mm}
  \centering
  \includegraphics[width=0.9\linewidth]{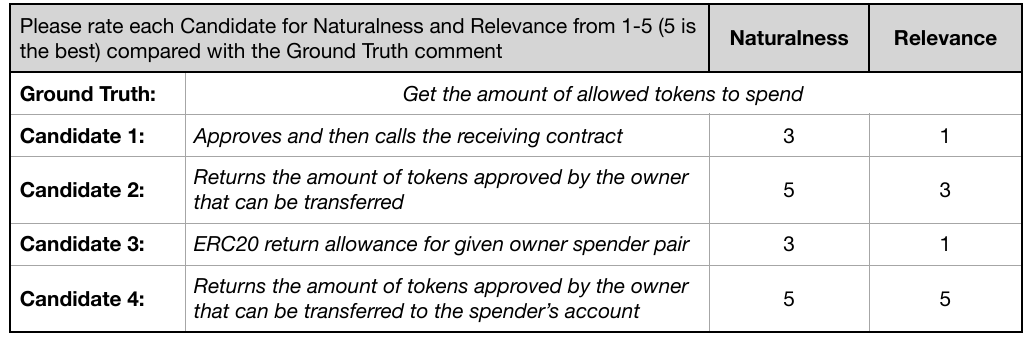} 
  \caption{A User Study Case} 
  \label{fig:case}
\end{figure}

The BM25, CodeT5 and PLBART are selected as baselines for human evaluation. 
Particularly, we invited 20 with 1-3 years of blockchain or smart contract experience and good English proficiency to evaluate the generated comments in the form of a questionnaire. 
An example of the questionnaire is demonstrated in Figure~\ref{fig:case}. 
We randomly sampled 25 $\langle$generated comment, reference comment$\rangle$ pairs from our evaluation dataset. 
To ensure the quality of the manual analysis, we divided the 20 participants into five groups, each group of participants will only rate 5 data samples independently. 
In other words, each data sample was rated by 4 independent evaluators. 
Each participant is required to compare the candidate comments with the ground truth comment and estimate their \textbf{Naturalness} and \textbf{Relevance} on a scale between 1 and 5 (5 is the best). 
During the annotation, participants are allowed to search the Internet for related information and unfamiliar concepts. 
Participants do not know which comment was generated by which approach.

\begin{table}[ht] 
  \caption{Human Evaluation} 
  \label{tab:user}
    \resizebox{\linewidth}{!}{
        \begin{tabular}{l|cccc|cccc}
        \toprule
        Method&N&$Low_{N}$&$Medium_{N}$&$High_{N}$&R&$Low_{R}$&$Medium_{R}$&$High_{R}$\\
        \midrule
        BM-25&3.57&15\%&31\%&54\%&1.74&85\%&6\%&9\%\\
        CodeT5&4.15&25\%&22\%&53\%&3.54&18\%&22\%&60\%\\
        PLBART&3.66&18\%&22\%&60\%&2.69&52\%&19\%&29\%\\
        \midrule
        \appnamebold&\textbf{4.34}&\textbf{0\%}&\textbf{12\%}&\textbf{88\%}&\textbf{3.72}&\textbf{18\%}&\textbf{18\%}&\textbf{64\%}\\
        \bottomrule
        \end{tabular}
    }
\end{table}

\subsubsection{Evaluation Results.}
After collecting responses from all evaluators, we regard a score of 1 and 2 as \textit{low quality}, a score of 3 as \textit{medium quality}, and a score of 4 and 5 as \textit{high quality}, respectively. 
We then calculated the average scores and the proportion of each quality type (e.g., \textit{low quality}, \textit{medium quality} and \textit{high quality}) respectively. 
The quality distribution and average score of naturalness and relevance across each method are presented in Table~\ref{tab:user}. 
From the table we can draw the following conclusions:
\begin{itemize}
    \item[(1)] \textbf{Regarding the naturalness score, \appnamebold outperforms other methods by a large margin}. 
    Notably, the proportion of low-quality comments is 0\%, while the proportion of high-quality comments is 88\%. 
    This indicates that our approach excels in generating fluent and grammatically correct comments. 
    \item[(2)] \textbf{Regarding the relevance score, \appnamebold also surpasses the other methods, indicating that the generated comments exhibit a higher degree of relevance to the comments written by developers.} 
    For instance, the average relevance score of \appname is 3.72, which is the highest among all other methods. 
    Furthermore, 64\% of the generated comments are rated of high quality by participants. 
    An example of this is demonstrated in Figure~\ref{fig:case}, candidate 4 (``\textit{returns the number of tokens approved by the owner that can be transferred to the spender's account}'') is the comment generated by our approach. 
    Even the generated comment and human written comment do not share many common words, but they are semantically equivalent and rated as high relevance scores by evaluators. 
    \item[(3)] \textbf{In general, our model performs well across both dimensions.} 
    The results of human evaluation are consistent with automatic evaluation results. 
    The considerable proportion of high-quality comments generated by our approach with respect to the \textbf{Naturalness} and \textbf{Relevance} also reconfirms the effectiveness of our system. 
\end{itemize}

\find{
\textbf{Answer to RQ-7: How effective is our \appname under human evaluation?} 
We conclude that under human evaluation, our approach outperforms other baselines in terms of both \textbf{Naturalness} and \textbf{Relevance} aspects. 
}



\section{THREATS TO VALIDITY}
\label{threats}
In this section, we discuss potential threats to the validity of our study, which are concerns related to the accuracy and generalizability of our findings. 

\noindent\textbf{Internal Validity:} 
Threats to interval validity primarily are concerned with potential errors that may have occurred in our code implementation and research settings. 
To minimize such errors, we have conducted thorough inspections and fully tested our source code in both model design and automatic evaluation. 
The parameters of the baseline methods have been carefully tuned and their highest-performing configurations are reported for comparison. 
However, despite these efforts, there remains a possibility of unnoticed errors in our implementation. 
Considering such cases, we have published our source code and dataset to facilitate other researchers to replicate and extend our work. 

\noindent\textbf{External validity:} 
Threats to external validity relate to the quality and generalizability of our dataset. 
Our dataset includes Solidity source code and their complied bytecode for smart contracts. 
There are other programming languages for smart contract development which are not considered in our study, we believe our results can generalize to other smart contract programming languages due to overall similarity in EVM bytecode representation. 
Future research will explore other smart contract languages' impact on code comment generation to enhance \appname's applicability.
Considering smart contracts have a very high clone ratio on the Ethereum blockchain, we performed data deduplication in our data preprocessing. 
In particular, we deduplicated our dataset according to the unique hash value generated by a contract's CFG and its comment. 
In other words, we can make sure our training set doesn't overlap with our testing set. 
A more comprehensive evaluation setting is to split the data timewisely (e.g., using historical data for training and future data for testing), we will explore the timewise evaluation of our approach in our future work.


\noindent\textbf{Construct validity:} 
Threats to construct validity relate to issues that could affect the ability to draw correct conclusions about the relations between the treatment and outcome of an experiment. 
Evaluation metrics suitability is the primary issue. 
We used BLEU and ROUGE for automatic evaluation, which are widely used for evaluating grammatical correctness and relevance. 
However, semantic understanding might be limited due to diverse natural language expressions. 
Although we incorporate manual evaluation, it can be affected by evaluator attentiveness, language proficiency, and blockchain expertise. 
To address this, we select experienced participants proficient in English and blockchain and provide each participant with reasonable data samples and sufficient time for evaluation. 
\section{DISCUSSION}
In this section, we discuss the adaptability of \appname to generate comments for other programming languages and the unique ability of \appname to generate comments for smart contracts. 

\subsection{Comment Generation from Bytecode}
\appname is proposed for handling smart contracts where their source code is missing, under such situations, only bytecode is available on the Ethereum blockchain. 
Therefore, we design \appname to directly translate smart contract bytecode to natural language comments. 
\textbf{Notably, even though our approach is designed for smart contracts, it can be easily extended to other programming languages}. 
In general, \appname can be regarded as a general framework for generating comments from bytecode, which uses CFG to capture the structural information and the IR-augmented component to capture semantic information. 
It is an interesting research direction to explore the effectiveness of \appname on other programming languages (e.g., Java), but it is beyond the scope of this research. 
Generating comments from bytecode not only benefits smart contract developers but also provides practical implications for other programming languages. 
For example, when developers face the task of decompilation (e.g., Android apps with only bytecode), the comments of the bytecode can provide a bird's eye view of the system's functionalities.

Recently, researchers have investigated of generating code comments from bytecode for popular programming languages, such as Java. 
For example, Huang et al.~\cite{huang2023bcgen} first proposed a method named BCGen to generate comments for Java bytecode in 2022. 
Similarly, they converted the bytecode into CFGs and built the neural language model to learn from the CFGs and token sequences. 
Our \appname differs from the existing studies of generating code summaries from bytecode for general languages in terms of the following aspects: 
1) Our research focuses on the EVM bytecode, this is because Solidity is specifically designed for writing smart contracts and is most widely used by smart contract developers. 
As far as we know, there is no ready-made dataset available for smart contracts and EVM bytecode, our work builds the first large dataset for this task. 
2) Previous studies find that smart contracts have a relatively high clone ratio (e.g., 90\%), much higher than the code clone ratio in traditional software projects~\cite{gao2020checking}. 
Inspired by this finding, we introduced the IR augmented component to retrieve the similar function's comment to assist the target comment generation. 
The experimental results verify the effectiveness of using IR components to enhance the overall performance of our approach.

\subsection{Comment Generation for Smart Contract}

The comment generation task has been widely explored by software engineering researchers, however, there are only a few studies that investigated comment generation task for smart contracts~\cite{hu2021automating, shi2023machine, yang2022ccgir}. 
Among them, two representative tools are closely related to our work, i.e., SmartDoc~\cite{hu2021automating} and Stan~\cite{li2020stan}. 
Hu et al.~\cite{hu2021automating} first introduce the task of generating descriptions for smart contracts from source code, they propose SmartDoc to generate user notices for smart contract functions. 
Our \appname shares the same architecture with SmartDoc by using the sequence-to-sequence learning of neural network models. 
However, different from SmartDoc which targets on smart contract source code, our approach first focuses on smart contract bytecode. 
However, only 13\% of smart contract source code are available, which means SmartDoc can only be applied to a small number of smart contracts, while our \appname can support all smart contracts because their bytecode are all available on-chain. 
Moreover, compared with the source code, the gap between comments and natural language comments is even larger, therefore we introduced the IR augmented component to bridge this semantic gap.

Li et al.~\cite{li2020stan} proposed a model named Stan to generate descriptions for bytecode of smart contracts. 
Compared with Stan, \appname is more general. 
Stan describes every smart contract interface from four aspects (i.e., functionality description, usage description, behavior description, and payment description). 
\appname aims to generate comments for smart contract functions, since comments are natural language descriptions written by developers, the comments generated by our approach are more general to describe the smart contract. 
Moreover, Stan requires to analyze the extra metadata and adopts symbolic execution techniques to generate intermediate information, while \appname is an end-to-end model that only acquires the runtime bytecode of a smart contract as input and automatically outputs the natural language comment. 
Therefore \appname is more lightweight and flexible compared with Stan.

\section{RELATED WORK}
\label{relatedwork}
We employ the deep learning model to generate natural language comments for the bytecode of the smart contract, which are mainly related to the following three main aspects.  

\subsection{Code Comment Generation}
Code comment generation is the most relevant task which aims to generate natural descriptions for code snippets. 
The Code comment generation task has been studied by a lot of software engineering researchers. 
Manually-crafted templates~\cite{moreno2013automatic,sridhara2010towards,mcburney2015automatic}, IR techniques~\cite{kuhn2007semantic,haiduc2010use,wong2013autocomment,wong2015clocom}, and neural network models~\cite{hu2018deep,wei2020retrieve,alon2018code2seq,wan2018improving} are widely used in automatic comment generation.
For example, Sridhara et al.~\cite{sridhara2010towards} propose to construct Software Word Usage Model (SWUM) to select relevant keywords from source code and then leverage them to construct natural language descriptions from defined templates. 
Haiduc et al.~\cite{haiduc2010use} exploit two IR techniques, Vector Space Model (VSM) and LSI, to analyze methods and classes in Java projects and generate short descriptions for them. 
Iyer et al.~\cite{iyer2016summarizing} first propose to utilize the encoder-decoder framework to generate comments, in which the encoder is token embeddings of source code and the decoder is an LSTM.
The experimental results on C\# and SQL comment generation illustrate that neural networks perform better than traditional techniques.

Despite the availability of various automated comment generation studies and  tools~\cite{gu2020codebert,yin2019codesummarization,allamanis2016codenn}, there are few tools specifically designed for generating comments on smart contract code~\cite{hu2021automating,shi2023machine,yang2022ccgir}. 
Hu et al.~\cite{hu2021automating} first introduce the task of generating descriptions for smart contracts from source code. 
However, generating smart contract comments from EVM bytecode has never been investigated. 
To the best of our knowledge, our work is the first research to explore the possibility of generating smart contract comments from EVM bytecode, and our extensive evaluation shows the effectiveness of our tool for this newly proposed task. 

\subsection{Smart Contract Bytecode Analysis}
Smart contract bytecode has been investigated in various tasks, including smart contract vulnerability detection~\cite{chen2021defectchecker, ashizawa2021eth2vec}, smart contract classification~\cite{shi2022bytecode, sezer2020exploiting}, and code similarity detection~\cite{zhu2022bytecode}. 
For example, Chen et al.~\cite{chen2021defectchecker} proposed DefectChecker, a symbolic execution-based approach and tool to detect eight contract defects that can cause unwanted behaviors of smart contracts on the Ethereum blockchain platform. DefectChecker can detect contract defects from smart contracts’ bytecode. 
Ashizawa et al.~\cite{ashizawa2021eth2vec} proposed Eth2Vec, a machine-learning-based static analysis tool for vulnerability detection in smart contracts. 
Eth2Vec automatically learns features of vulnerable EVM bytecodes with tacit knowledge through a neural network for natural language processing. 
Shi et al.~\cite{shi2022bytecode} proposed a novel bytecode-based classification
approach to effectively classify smart contracts of blockchain platforms. 
There are also techniques investigating how to decompile smart contract bytecode and generate Solidity source code. 
For example, Grech et al~\cite{grech2019gigahorse} presented Gigahorse, which is a reverse compiler that decompiles the smart contract EVM bytecode into high-level three-address code representation. Following that, they proposed Elipmoc~\cite{grech2022elipmoc} to further improve Gigahorse by integrating high-precision algorithms and design decisions that target a balance of precision and scalability. 
Suiche et al.~\cite{suiche2017porosity} proposed Porosity, which is able to generate readable Solidity syntax contracts and enable static or dynamic analysis on these compiled contracts. 
However, even with these decompilers, it is still not easy for users to grasp the semantic information of the contract, not to mention the potential misleading due to decompilation errors.

Different from previous studies that focus on smart contract vulnerability detection from bytecode, our work first introduces the task of generating descriptions for smart contracts from EVM bytecode. 
Considering more than 90\% of smart contract source codes are not available on the blockchain, while their EVM bytecode is always accessible, our tool complements the existing smart contract comment generation tools by using bytecode. 

\subsection{Software Engineering on Smart Contract}
Software engineering researchers have investigated smart contracts for different software engineering tasks, such as smart contract clone detection~\cite{gao2020checking, gao2019smartembed, kondo2020code, gao2020deep, chen2024angels}, smart contract code searching~\cite{shi2021semantic}, smart contract program repair~\cite{yu2020smart, wang2022empirical}. 
For example, Gao et al.~\cite{gao2020checking} proposed a model, named SmartEmbed to detect code clones and clone-related bugs in smart contracts by using structural code embedding techniques. 
Shi et al.~\cite{shi2021semantic} proposed a Multi-Modal Smart contract Code Search (MM-SCS) model for semantic code search with smart contracts. 
Their model can capture the data-flow and control-flow information from source code as well as semantic features. 
Yu et al.~\cite{yu2020smart} proposed the first general-purpose automated smart contract repair approach that is also gas-aware. 
Their program repair model is search-based and searches among mutations of the buggy contract. 
Compared with the aforementioned studies, our research focuses on smart contract code comprehension and maintenance, our approach can greatly benefit smart contract developers and end users in understanding the functionality and logic of smart contracts from EVM bytecode instead of source code.  

\section{CONCLUSION}
\label{conclusion}

In this work, we first introduce the task of generating smart contract function descriptions from their EVM bytecode. 
We have proposed a novel model, named \appname, for translating bytecode to function comment automatically. 
\appname employs IR augmented module to fill the semantic gap between bytecode and natural language comments and adopts the encoder-decoder neural network to learn structural information from smart contract bytecode. 
We have conducted extensive experiments to verify the effectiveness of our approach, and \appname gets remarkable automatic evaluation scores and understandable human evaluation. 

\section{Acknowledgments}
This research/project is supported by the National Key Research and Development Program of China (No. 2021YFB2701102), the National Science Foundation of China (No.62372398, No. 62202341, No.72342025, and U20A20173), the Fundamental Research Funds for the Central Universities (No. 226-2022-00064). 
This research is partially supported by the Shanghai Sailing Program (23YF1446900). 
This research is partially supported by the Starry Night Science Fund of Zhejiang University Shanghai Institute for Advanced Study, Grant No. SN-ZJU-SIAS-001. 
This research is partially supported by the Ningbo Natural Science Foundation (No. 2023J292). 
This research is also supported by the advanced computing resources provided by the Supercomputing Center of Hangzhou City University. 
The authors would like to thank the reviewers for their insightful and constructive feedback.

\bibliographystyle{ACM-Reference-Format}
\bibliography{document}


\begin{thebibliography}{71}


\ifx \showCODEN    \undefined \def \showCODEN     #1{\unskip}     \fi
\ifx \showDOI      \undefined \def \showDOI       #1{#1}\fi
\ifx \showISBNx    \undefined \def \showISBNx     #1{\unskip}     \fi
\ifx \showISBNxiii \undefined \def \showISBNxiii  #1{\unskip}     \fi
\ifx \showISSN     \undefined \def \showISSN      #1{\unskip}     \fi
\ifx \showLCCN     \undefined \def \showLCCN      #1{\unskip}     \fi
\ifx \shownote     \undefined \def \shownote      #1{#1}          \fi
\ifx \showarticletitle \undefined \def \showarticletitle #1{#1}   \fi
\ifx \showURL      \undefined \def \showURL       {\relax}        \fi
\providecommand\bibfield[2]{#2}
\providecommand\bibinfo[2]{#2}
\providecommand\natexlab[1]{#1}
\providecommand\showeprint[2][]{arXiv:#2}

\bibitem[Bit(2018)]%
        {Bitcoinafrica}
 \bibinfo{year}{2018}\natexlab{}.
\newblock \bibinfo{booktitle}{\emph{The Biggest {ICO} Scams of 2018}}.
\newblock
\urldef\tempurl%
\url{https://bitcoinafrica.io/2018/09/27/biggest-ico-scams/}
\showURL{%
Retrieved July 10, 2023 from \tempurl}


\bibitem[Evm(2019)]%
        {EvmCfgBuilder}
 \bibinfo{year}{2019}\natexlab{}.
\newblock \bibinfo{booktitle}{\emph{EVM CFG Builder GitHub Repository}}.
\newblock
\urldef\tempurl%
\url{https://github.com/crytic/evm_cfg_builder}
\showURL{%
Retrieved July 10, 2023 from \tempurl}


\bibitem[Eth(2023)]%
        {Etherscan}
 \bibinfo{year}{2023}\natexlab{}.
\newblock \bibinfo{booktitle}{\emph{Etherscan Ethereum Blockchain Explorer}}.
\newblock
\urldef\tempurl%
\url{https://etherscan.io/}
\showURL{%
Retrieved July 10, 2023 from \tempurl}


\bibitem[Zel(2023)]%
        {Zellic}
 \bibinfo{year}{2023}\natexlab{}.
\newblock \bibinfo{booktitle}{\emph{Zellic}}.
\newblock
\urldef\tempurl%
\url{https://www.zellic.io/}
\showURL{%
Retrieved July 10, 2023 from \tempurl}


\bibitem[Ahmad et~al\mbox{.}(2021)]%
        {ahmad2021unified}
\bibfield{author}{\bibinfo{person}{Wasi~Uddin Ahmad}, \bibinfo{person}{Saikat Chakraborty}, \bibinfo{person}{Baishakhi Ray}, {and} \bibinfo{person}{Kai-Wei Chang}.} \bibinfo{year}{2021}\natexlab{}.
\newblock \showarticletitle{Unified pre-training for program understanding and generation}.
\newblock \bibinfo{journal}{\emph{arXiv preprint arXiv:2103.06333}} (\bibinfo{year}{2021}).
\newblock


\bibitem[Allamanis et~al\mbox{.}(2016)]%
        {allamanis2016codenn}
\bibfield{author}{\bibinfo{person}{Miltiadis Allamanis}, \bibinfo{person}{Hao Peng}, {and} \bibinfo{person}{Charles Sutton}.} \bibinfo{year}{2016}\natexlab{}.
\newblock \showarticletitle{CodeNN: Neural Code Representation Learning for Code Summarization}. In \bibinfo{booktitle}{\emph{Proceedings of the 40th International Conference on Software Engineering (ICSE)}}.
\newblock


\bibitem[Alon et~al\mbox{.}(2018)]%
        {alon2018code2seq}
\bibfield{author}{\bibinfo{person}{Uri Alon}, \bibinfo{person}{Shaked Brody}, \bibinfo{person}{Omer Levy}, {and} \bibinfo{person}{Eran Yahav}.} \bibinfo{year}{2018}\natexlab{}.
\newblock \showarticletitle{code2seq: Generating Sequences from Structured Representations of Code}. In \bibinfo{booktitle}{\emph{International Conference on Learning Representations}}.
\newblock


\bibitem[{Anonymous}(2023)]%
        {smartbt_dataset}
\bibfield{author}{\bibinfo{person}{{Anonymous}}.} \bibinfo{year}{2023}\natexlab{}.
\newblock \bibinfo{title}{SmartBT Dataset}.
\newblock
\newblock
\urldef\tempurl%
\url{https://figshare.com/s/e5f3f9371ca67a63c122}
\showURL{%
Retrieved July 28, 2023 from \tempurl}


\bibitem[Ashizawa et~al\mbox{.}(2021)]%
        {ashizawa2021eth2vec}
\bibfield{author}{\bibinfo{person}{Nami Ashizawa}, \bibinfo{person}{Naoto Yanai}, \bibinfo{person}{Jason~Paul Cruz}, {and} \bibinfo{person}{Shingo Okamura}.} \bibinfo{year}{2021}\natexlab{}.
\newblock \showarticletitle{Eth2Vec: learning contract-wide code representations for vulnerability detection on ethereum smart contracts}. In \bibinfo{booktitle}{\emph{Proceedings of the 3rd ACM international symposium on blockchain and secure critical infrastructure}}. \bibinfo{pages}{47--59}.
\newblock


\bibitem[Bahdanau et~al\mbox{.}(2014)]%
        {bahdanau2014neural}
\bibfield{author}{\bibinfo{person}{Dzmitry Bahdanau}, \bibinfo{person}{Kyunghyun Cho}, {and} \bibinfo{person}{Yoshua Bengio}.} \bibinfo{year}{2014}\natexlab{}.
\newblock \showarticletitle{Neural machine translation by jointly learning to align and translate}.
\newblock \bibinfo{journal}{\emph{arXiv preprint arXiv:1409.0473}} (\bibinfo{year}{2014}).
\newblock


\bibitem[Chen et~al\mbox{.}(2024)]%
        {chen2024angels}
\bibfield{author}{\bibinfo{person}{Jiachi Chen}, \bibinfo{person}{Jiang Hu}, \bibinfo{person}{Xin Xia}, \bibinfo{person}{David Lo}, \bibinfo{person}{John Grundy}, \bibinfo{person}{Zhipeng Gao}, {and} \bibinfo{person}{Ting Chen}.} \bibinfo{year}{2024}\natexlab{}.
\newblock \showarticletitle{Angels or demons: investigating and detecting decentralized financial traps on ethereum smart contracts}.
\newblock \bibinfo{journal}{\emph{Automated Software Engineering}} \bibinfo{volume}{31}, \bibinfo{number}{2} (\bibinfo{year}{2024}), \bibinfo{pages}{63}.
\newblock


\bibitem[Chen et~al\mbox{.}(2021a)]%
        {chen2021smart}
\bibfield{author}{\bibinfo{person}{Jiachi Chen}, \bibinfo{person}{Xin Xia}, \bibinfo{person}{David Lo}, {and} \bibinfo{person}{John Grundy}.} \bibinfo{year}{2021}\natexlab{a}.
\newblock \showarticletitle{Why do smart contracts self-destruct? investigating the selfdestruct function on ethereum}.
\newblock \bibinfo{journal}{\emph{ACM Transactions on Software Engineering and Methodology (TOSEM)}} \bibinfo{volume}{31}, \bibinfo{number}{2} (\bibinfo{year}{2021}), \bibinfo{pages}{1--37}.
\newblock


\bibitem[Chen et~al\mbox{.}(2021b)]%
        {chen2021defectchecker}
\bibfield{author}{\bibinfo{person}{Jiachi Chen}, \bibinfo{person}{Xin Xia}, \bibinfo{person}{David Lo}, \bibinfo{person}{John Grundy}, \bibinfo{person}{Xiapu Luo}, {and} \bibinfo{person}{Ting Chen}.} \bibinfo{year}{2021}\natexlab{b}.
\newblock \showarticletitle{Defectchecker: Automated smart contract defect detection by analyzing evm bytecode}.
\newblock \bibinfo{journal}{\emph{IEEE Transactions on Software Engineering}} \bibinfo{volume}{48}, \bibinfo{number}{7} (\bibinfo{year}{2021}), \bibinfo{pages}{2189--2207}.
\newblock


\bibitem[Dai et~al\mbox{.}(2024)]%
        {dai2024mpcoder}
\bibfield{author}{\bibinfo{person}{Zhenlong Dai}, \bibinfo{person}{Chang Yao}, \bibinfo{person}{WenKang Han}, \bibinfo{person}{Yuanying Yuanying}, \bibinfo{person}{Zhipeng Gao}, {and} \bibinfo{person}{Jingyuan Chen}.} \bibinfo{year}{2024}\natexlab{}.
\newblock \showarticletitle{MPCoder: Multi-user Personalized Code Generator with Explicit and Implicit Style Representation Learning}. In \bibinfo{booktitle}{\emph{Proceedings of the 62nd Annual Meeting of the Association for Computational Linguistics (Volume 1: Long Papers)}}. \bibinfo{pages}{3765--3780}.
\newblock


\bibitem[Gao(2020)]%
        {gao2020deep}
\bibfield{author}{\bibinfo{person}{Zhipeng Gao}.} \bibinfo{year}{2020}\natexlab{}.
\newblock \showarticletitle{When deep learning meets smart contracts}. In \bibinfo{booktitle}{\emph{Proceedings of the 35th IEEE/ACM International Conference on Automated Software Engineering}}. \bibinfo{pages}{1400--1402}.
\newblock


\bibitem[Gao et~al\mbox{.}(2019)]%
        {gao2019smartembed}
\bibfield{author}{\bibinfo{person}{Zhipeng Gao}, \bibinfo{person}{Vinoj Jayasundara}, \bibinfo{person}{Lingxiao Jiang}, \bibinfo{person}{Xin Xia}, \bibinfo{person}{David Lo}, {and} \bibinfo{person}{John Grundy}.} \bibinfo{year}{2019}\natexlab{}.
\newblock \showarticletitle{Smartembed: A tool for clone and bug detection in smart contracts through structural code embedding}. In \bibinfo{booktitle}{\emph{2019 IEEE International Conference on Software Maintenance and Evolution (ICSME)}}. IEEE, \bibinfo{pages}{394--397}.
\newblock


\bibitem[Gao et~al\mbox{.}(2020a)]%
        {gao2020checking}
\bibfield{author}{\bibinfo{person}{Zhipeng Gao}, \bibinfo{person}{Lingxiao Jiang}, \bibinfo{person}{Xin Xia}, \bibinfo{person}{David Lo}, {and} \bibinfo{person}{John Grundy}.} \bibinfo{year}{2020}\natexlab{a}.
\newblock \showarticletitle{Checking smart contracts with structural code embedding}.
\newblock \bibinfo{journal}{\emph{IEEE Transactions on Software Engineering}} \bibinfo{volume}{47}, \bibinfo{number}{12} (\bibinfo{year}{2020}), \bibinfo{pages}{2874--2891}.
\newblock


\bibitem[Gao et~al\mbox{.}(2020b)]%
        {gao2020generating}
\bibfield{author}{\bibinfo{person}{Zhipeng Gao}, \bibinfo{person}{Xin Xia}, \bibinfo{person}{John Grundy}, \bibinfo{person}{David Lo}, {and} \bibinfo{person}{Yuan-Fang Li}.} \bibinfo{year}{2020}\natexlab{b}.
\newblock \showarticletitle{Generating question titles for stack overflow from mined code snippets}.
\newblock \bibinfo{journal}{\emph{ACM Transactions on Software Engineering and Methodology (TOSEM)}} \bibinfo{volume}{29}, \bibinfo{number}{4} (\bibinfo{year}{2020}), \bibinfo{pages}{1--37}.
\newblock


\bibitem[Gao et~al\mbox{.}(2020c)]%
        {gao2020technical}
\bibfield{author}{\bibinfo{person}{Zhipeng Gao}, \bibinfo{person}{Xin Xia}, \bibinfo{person}{David Lo}, {and} \bibinfo{person}{John Grundy}.} \bibinfo{year}{2020}\natexlab{c}.
\newblock \showarticletitle{Technical Q8A site answer recommendation via question boosting}.
\newblock \bibinfo{journal}{\emph{ACM Transactions on Software Engineering and Methodology (TOSEM)}} \bibinfo{volume}{30}, \bibinfo{number}{1} (\bibinfo{year}{2020}), \bibinfo{pages}{1--34}.
\newblock


\bibitem[Gao et~al\mbox{.}(2023)]%
        {gao2023know}
\bibfield{author}{\bibinfo{person}{Zhipeng Gao}, \bibinfo{person}{Xin Xia}, \bibinfo{person}{David Lo}, \bibinfo{person}{John Grundy}, \bibinfo{person}{Xindong Zhang}, {and} \bibinfo{person}{Zhenchang Xing}.} \bibinfo{year}{2023}\natexlab{}.
\newblock \showarticletitle{I know what you are searching for: Code snippet recommendation from stack overflow posts}.
\newblock \bibinfo{journal}{\emph{ACM Transactions on Software Engineering and Methodology}} \bibinfo{volume}{32}, \bibinfo{number}{3} (\bibinfo{year}{2023}), \bibinfo{pages}{1--42}.
\newblock


\bibitem[Grech et~al\mbox{.}(2019)]%
        {grech2019gigahorse}
\bibfield{author}{\bibinfo{person}{Neville Grech}, \bibinfo{person}{Lexi Brent}, \bibinfo{person}{Bernhard Scholz}, {and} \bibinfo{person}{Yannis Smaragdakis}.} \bibinfo{year}{2019}\natexlab{}.
\newblock \showarticletitle{Gigahorse: thorough, declarative decompilation of smart contracts}. In \bibinfo{booktitle}{\emph{2019 IEEE/ACM 41st International Conference on Software Engineering (ICSE)}}. IEEE, \bibinfo{pages}{1176--1186}.
\newblock


\bibitem[Grech et~al\mbox{.}(2022)]%
        {grech2022elipmoc}
\bibfield{author}{\bibinfo{person}{Neville Grech}, \bibinfo{person}{Sifis Lagouvardos}, \bibinfo{person}{Ilias Tsatiris}, {and} \bibinfo{person}{Yannis Smaragdakis}.} \bibinfo{year}{2022}\natexlab{}.
\newblock \showarticletitle{Elipmoc: advanced decompilation of ethereum smart contracts}.
\newblock \bibinfo{journal}{\emph{Proceedings of the ACM on Programming Languages}} \bibinfo{volume}{6}, \bibinfo{number}{OOPSLA1} (\bibinfo{year}{2022}), \bibinfo{pages}{1--27}.
\newblock


\bibitem[Gu et~al\mbox{.}(2016a)]%
        {gu2016incorporating}
\bibfield{author}{\bibinfo{person}{Jiatao Gu}, \bibinfo{person}{Zhengdong Lu}, \bibinfo{person}{Hang Li}, {and} \bibinfo{person}{Victor~OK Li}.} \bibinfo{year}{2016}\natexlab{a}.
\newblock \showarticletitle{Incorporating copying mechanism in sequence-to-sequence learning}.
\newblock \bibinfo{journal}{\emph{arXiv preprint arXiv:1603.06393}} (\bibinfo{year}{2016}).
\newblock


\bibitem[Gu et~al\mbox{.}(2020)]%
        {gu2020codebert}
\bibfield{author}{\bibinfo{person}{Xiaodong Gu}, \bibinfo{person}{Kelvin Guu}, {and} \bibinfo{person}{Geoffrey Hinton}.} \bibinfo{year}{2020}\natexlab{}.
\newblock \showarticletitle{CodeBERT: A Pre-trained Model for Programming and Natural Languages}. In \bibinfo{booktitle}{\emph{Conference on Empirical Methods in Natural Language Processing (EMNLP)}}.
\newblock


\bibitem[Gu et~al\mbox{.}(2016b)]%
        {gu2016deep}
\bibfield{author}{\bibinfo{person}{Xiaodong Gu}, \bibinfo{person}{Hongyu Zhang}, \bibinfo{person}{Dongmei Zhang}, {and} \bibinfo{person}{Sunghun Kim}.} \bibinfo{year}{2016}\natexlab{b}.
\newblock \showarticletitle{Deep API learning}. In \bibinfo{booktitle}{\emph{Proceedings of the 2016 24th ACM SIGSOFT International Symposium on Foundations of Software Engineering}}. \bibinfo{pages}{631--642}.
\newblock


\bibitem[Guo et~al\mbox{.}(2020)]%
        {guo2020graphcodebert}
\bibfield{author}{\bibinfo{person}{Daya Guo}, \bibinfo{person}{Shuo Ren}, \bibinfo{person}{Shuai Lu}, \bibinfo{person}{Zhangyin Feng}, \bibinfo{person}{Duyu Tang}, \bibinfo{person}{Shujie Liu}, \bibinfo{person}{Long Zhou}, \bibinfo{person}{Nan Duan}, \bibinfo{person}{Alexey Svyatkovskiy}, \bibinfo{person}{Shengyu Fu}, {et~al\mbox{.}}} \bibinfo{year}{2020}\natexlab{}.
\newblock \showarticletitle{Graphcodebert: Pre-training code representations with data flow}.
\newblock \bibinfo{journal}{\emph{arXiv preprint arXiv:2009.08366}} (\bibinfo{year}{2020}).
\newblock


\bibitem[Guo et~al\mbox{.}(2023)]%
        {guo2023large}
\bibfield{author}{\bibinfo{person}{Taicheng Guo}, \bibinfo{person}{Kehan Guo}, \bibinfo{person}{Bozhao Nan}, \bibinfo{person}{Zhenwen Liang}, \bibinfo{person}{Zhichun Guo}, \bibinfo{person}{Nitesh~V. Chawla}, \bibinfo{person}{Olaf Wiest}, {and} \bibinfo{person}{Xiangliang Zhang}.} \bibinfo{year}{2023}\natexlab{}.
\newblock \bibinfo{title}{What can Large Language Models do in chemistry? A comprehensive benchmark on eight tasks}.
\newblock
\newblock
\showeprint[arxiv]{2305.18365}~[cs.CL]


\bibitem[Haiduc et~al\mbox{.}(2010)]%
        {haiduc2010use}
\bibfield{author}{\bibinfo{person}{Sonia Haiduc}, \bibinfo{person}{Jairo Aponte}, \bibinfo{person}{Laura Moreno}, {and} \bibinfo{person}{Andrian Marcus}.} \bibinfo{year}{2010}\natexlab{}.
\newblock \showarticletitle{On the use of automated text summarization techniques for summarizing source code}. In \bibinfo{booktitle}{\emph{2010 17th Working Conference on Reverse Engineering}}. IEEE, \bibinfo{pages}{35--44}.
\newblock


\bibitem[Hu et~al\mbox{.}(2021)]%
        {hu2021automating}
\bibfield{author}{\bibinfo{person}{Xing Hu}, \bibinfo{person}{Zhipeng Gao}, \bibinfo{person}{Xin Xia}, \bibinfo{person}{David Lo}, {and} \bibinfo{person}{Xiaohu Yang}.} \bibinfo{year}{2021}\natexlab{}.
\newblock \showarticletitle{Automating user notice generation for smart contract functions}. In \bibinfo{booktitle}{\emph{2021 36th IEEE/ACM International Conference on Automated Software Engineering (ASE)}}. IEEE, \bibinfo{pages}{5--17}.
\newblock


\bibitem[Hu et~al\mbox{.}(2018)]%
        {hu2018deep}
\bibfield{author}{\bibinfo{person}{Xing Hu}, \bibinfo{person}{Ge Li}, \bibinfo{person}{Xin Xia}, \bibinfo{person}{David Lo}, {and} \bibinfo{person}{Zhi Jin}.} \bibinfo{year}{2018}\natexlab{}.
\newblock \showarticletitle{Deep code comment generation}. In \bibinfo{booktitle}{\emph{2018 IEEE/ACM 26th International Conference on Program Comprehension (ICPC)}}. IEEE, \bibinfo{pages}{200--20010}.
\newblock


\bibitem[Hu et~al\mbox{.}(2020)]%
        {hu2020deep}
\bibfield{author}{\bibinfo{person}{Xing Hu}, \bibinfo{person}{Ge Li}, \bibinfo{person}{Xin Xia}, \bibinfo{person}{David Lo}, {and} \bibinfo{person}{Zhi Jin}.} \bibinfo{year}{2020}\natexlab{}.
\newblock \showarticletitle{Deep code comment generation with hybrid lexical and syntactical information}.
\newblock \bibinfo{journal}{\emph{Empirical Software Engineering}} \bibinfo{volume}{25}, \bibinfo{number}{3} (\bibinfo{year}{2020}), \bibinfo{pages}{2179--2217}.
\newblock


\bibitem[Huang et~al\mbox{.}(2023)]%
        {huang2023bcgen}
\bibfield{author}{\bibinfo{person}{Yuan Huang}, \bibinfo{person}{Jinbo Huang}, \bibinfo{person}{Xiangping Chen}, \bibinfo{person}{Kunning He}, {and} \bibinfo{person}{Xiaocong Zhou}.} \bibinfo{year}{2023}\natexlab{}.
\newblock \showarticletitle{BCGen: a comment generation method for bytecode}.
\newblock \bibinfo{journal}{\emph{Automated Software Engineering}} \bibinfo{volume}{30}, \bibinfo{number}{1} (\bibinfo{year}{2023}), \bibinfo{pages}{5}.
\newblock


\bibitem[Iyer et~al\mbox{.}(2016)]%
        {iyer2016summarizing}
\bibfield{author}{\bibinfo{person}{Srinivasan Iyer}, \bibinfo{person}{Ioannis Konstas}, \bibinfo{person}{Alvin Cheung}, {and} \bibinfo{person}{Luke Zettlemoyer}.} \bibinfo{year}{2016}\natexlab{}.
\newblock \showarticletitle{Summarizing source code using a neural attention model}. In \bibinfo{booktitle}{\emph{54th Annual Meeting of the Association for Computational Linguistics 2016}}. Association for Computational Linguistics, \bibinfo{pages}{2073--2083}.
\newblock


\bibitem[Jiang et~al\mbox{.}(2017)]%
        {jiang2017automatically}
\bibfield{author}{\bibinfo{person}{Siyuan Jiang}, \bibinfo{person}{Ameer Armaly}, {and} \bibinfo{person}{Collin McMillan}.} \bibinfo{year}{2017}\natexlab{}.
\newblock \showarticletitle{Automatically generating commit messages from diffs using neural machine translation}. In \bibinfo{booktitle}{\emph{2017 32nd IEEE/ACM International Conference on Automated Software Engineering (ASE)}}. IEEE, \bibinfo{pages}{135--146}.
\newblock


\bibitem[Kingma and Ba(2014)]%
        {kingma2014adam}
\bibfield{author}{\bibinfo{person}{Diederik~P Kingma} {and} \bibinfo{person}{Jimmy Ba}.} \bibinfo{year}{2014}\natexlab{}.
\newblock \showarticletitle{Adam: A method for stochastic optimization}.
\newblock \bibinfo{journal}{\emph{arXiv preprint arXiv:1412.6980}} (\bibinfo{year}{2014}).
\newblock


\bibitem[Kondo et~al\mbox{.}(2020)]%
        {kondo2020code}
\bibfield{author}{\bibinfo{person}{Masanari Kondo}, \bibinfo{person}{Gustavo~A Oliva}, \bibinfo{person}{Zhen~Ming Jiang}, \bibinfo{person}{Ahmed~E Hassan}, {and} \bibinfo{person}{Osamu Mizuno}.} \bibinfo{year}{2020}\natexlab{}.
\newblock \showarticletitle{Code cloning in smart contracts: a case study on verified contracts from the Ethereum blockchain platform}.
\newblock \bibinfo{journal}{\emph{Empirical Software Engineering}}  \bibinfo{volume}{25} (\bibinfo{year}{2020}), \bibinfo{pages}{4617--4675}.
\newblock


\bibitem[Kuhn et~al\mbox{.}(2007)]%
        {kuhn2007semantic}
\bibfield{author}{\bibinfo{person}{Adrian Kuhn}, \bibinfo{person}{St{\'e}phane Ducasse}, {and} \bibinfo{person}{Tudor G{\'\i}rba}.} \bibinfo{year}{2007}\natexlab{}.
\newblock \showarticletitle{Semantic clustering: Identifying topics in source code}.
\newblock \bibinfo{journal}{\emph{Information and software technology}} \bibinfo{volume}{49}, \bibinfo{number}{3} (\bibinfo{year}{2007}), \bibinfo{pages}{230--243}.
\newblock


\bibitem[Li et~al\mbox{.}(2020)]%
        {li2020stan}
\bibfield{author}{\bibinfo{person}{Xiaoqi Li}, \bibinfo{person}{Ting Chen}, \bibinfo{person}{Xiapu Luo}, \bibinfo{person}{Tao Zhang}, \bibinfo{person}{Le Yu}, {and} \bibinfo{person}{Zhou Xu}.} \bibinfo{year}{2020}\natexlab{}.
\newblock \showarticletitle{Stan: Towards describing bytecodes of smart contract}. In \bibinfo{booktitle}{\emph{2020 IEEE 20th International Conference on Software Quality, Reliability and Security (QRS)}}. IEEE, \bibinfo{pages}{273--284}.
\newblock


\bibitem[Lin(2004)]%
        {lin2004rouge}
\bibfield{author}{\bibinfo{person}{Chin-Yew Lin}.} \bibinfo{year}{2004}\natexlab{}.
\newblock \showarticletitle{Rouge: A package for automatic evaluation of summaries}. In \bibinfo{booktitle}{\emph{Text summarization branches out}}. \bibinfo{pages}{74--81}.
\newblock


\bibitem[Mai et~al\mbox{.}(2024)]%
        {mai2024human}
\bibfield{author}{\bibinfo{person}{Yubo Mai}, \bibinfo{person}{Zhipeng Gao}, \bibinfo{person}{Xing Hu}, \bibinfo{person}{Lingfeng Bao}, \bibinfo{person}{Yu Liu}, {and} \bibinfo{person}{JianLing Sun}.} \bibinfo{year}{2024}\natexlab{}.
\newblock \showarticletitle{Are Human Rules Necessary? Generating Reusable APIs with CoT Reasoning and In-Context Learning}.
\newblock \bibinfo{journal}{\emph{Proceedings of the ACM on Software Engineering}} \bibinfo{volume}{1}, \bibinfo{number}{FSE} (\bibinfo{year}{2024}), \bibinfo{pages}{2355--2377}.
\newblock


\bibitem[McBurney and McMillan(2015)]%
        {mcburney2015automatic}
\bibfield{author}{\bibinfo{person}{Paul~W McBurney} {and} \bibinfo{person}{Collin McMillan}.} \bibinfo{year}{2015}\natexlab{}.
\newblock \showarticletitle{Automatic source code summarization of context for java methods}.
\newblock \bibinfo{journal}{\emph{IEEE Transactions on Software Engineering}} \bibinfo{volume}{42}, \bibinfo{number}{2} (\bibinfo{year}{2015}), \bibinfo{pages}{103--119}.
\newblock


\bibitem[Moreno et~al\mbox{.}(2013)]%
        {moreno2013automatic}
\bibfield{author}{\bibinfo{person}{Laura Moreno}, \bibinfo{person}{Jairo Aponte}, \bibinfo{person}{Giriprasad Sridhara}, \bibinfo{person}{Andrian Marcus}, \bibinfo{person}{Lori Pollock}, {and} \bibinfo{person}{K Vijay-Shanker}.} \bibinfo{year}{2013}\natexlab{}.
\newblock \showarticletitle{Automatic generation of natural language summaries for java classes}. In \bibinfo{booktitle}{\emph{2013 21st International Conference on Program Comprehension (ICPC)}}. IEEE, \bibinfo{pages}{23--32}.
\newblock


\bibitem[Papineni et~al\mbox{.}(2002)]%
        {papineni2002bleu}
\bibfield{author}{\bibinfo{person}{Kishore Papineni}, \bibinfo{person}{Salim Roukos}, \bibinfo{person}{Todd Ward}, {and} \bibinfo{person}{Wei-Jing Zhu}.} \bibinfo{year}{2002}\natexlab{}.
\newblock \showarticletitle{Bleu: a method for automatic evaluation of machine translation}. In \bibinfo{booktitle}{\emph{Proceedings of the 40th annual meeting of the Association for Computational Linguistics}}. \bibinfo{pages}{311--318}.
\newblock


\bibitem[Qiu et~al\mbox{.}(2021)]%
        {qiu2021deep}
\bibfield{author}{\bibinfo{person}{Fangcheng Qiu}, \bibinfo{person}{Zhipeng Gao}, \bibinfo{person}{Xin Xia}, \bibinfo{person}{David Lo}, \bibinfo{person}{John Grundy}, {and} \bibinfo{person}{Xinyu Wang}.} \bibinfo{year}{2021}\natexlab{}.
\newblock \showarticletitle{Deep just-in-time defect localization}.
\newblock \bibinfo{journal}{\emph{IEEE Transactions on Software Engineering}} \bibinfo{volume}{48}, \bibinfo{number}{12} (\bibinfo{year}{2021}), \bibinfo{pages}{5068--5086}.
\newblock


\bibitem[Raffel et~al\mbox{.}(2020)]%
        {raffel2020exploring}
\bibfield{author}{\bibinfo{person}{Colin Raffel}, \bibinfo{person}{Noam Shazeer}, \bibinfo{person}{Adam Roberts}, \bibinfo{person}{Katherine Lee}, \bibinfo{person}{Sharan Narang}, \bibinfo{person}{Michael Matena}, \bibinfo{person}{Yanqi Zhou}, \bibinfo{person}{Wei Li}, {and} \bibinfo{person}{Peter~J Liu}.} \bibinfo{year}{2020}\natexlab{}.
\newblock \showarticletitle{Exploring the limits of transfer learning with a unified text-to-text transformer}.
\newblock \bibinfo{journal}{\emph{The Journal of Machine Learning Research}} \bibinfo{volume}{21}, \bibinfo{number}{1} (\bibinfo{year}{2020}), \bibinfo{pages}{5485--5551}.
\newblock


\bibitem[Robertson et~al\mbox{.}(2009)]%
        {robertson2009probabilistic}
\bibfield{author}{\bibinfo{person}{Stephen Robertson}, \bibinfo{person}{Hugo Zaragoza}, {et~al\mbox{.}}} \bibinfo{year}{2009}\natexlab{}.
\newblock \showarticletitle{The probabilistic relevance framework: BM25 and beyond}.
\newblock \bibinfo{journal}{\emph{Foundations and Trends{\textregistered} in Information Retrieval}} \bibinfo{volume}{3}, \bibinfo{number}{4} (\bibinfo{year}{2009}), \bibinfo{pages}{333--389}.
\newblock


\bibitem[Salton and Buckley(1988)]%
        {salton1988term}
\bibfield{author}{\bibinfo{person}{Gerard Salton} {and} \bibinfo{person}{Christopher Buckley}.} \bibinfo{year}{1988}\natexlab{}.
\newblock \showarticletitle{Term-weighting approaches in automatic text retrieval}.
\newblock \bibinfo{journal}{\emph{Information processing \& management}} \bibinfo{volume}{24}, \bibinfo{number}{5} (\bibinfo{year}{1988}), \bibinfo{pages}{513--523}.
\newblock


\bibitem[Sezer et~al\mbox{.}(2020)]%
        {sezer2020exploiting}
\bibfield{author}{\bibinfo{person}{Selin Sezer}, \bibinfo{person}{Clemens Eyhoff}, \bibinfo{person}{Wolfgang Prinz}, {and} \bibinfo{person}{Thomas Rose}.} \bibinfo{year}{2020}\natexlab{}.
\newblock \showarticletitle{Exploiting Smart Contract Bytecode for Classification on Ethereum.}. In \bibinfo{booktitle}{\emph{PoEM Workshops}}. \bibinfo{pages}{11--22}.
\newblock


\bibitem[Shi et~al\mbox{.}(2021)]%
        {shi2021semantic}
\bibfield{author}{\bibinfo{person}{Chaochen Shi}, \bibinfo{person}{Yong Xiang}, \bibinfo{person}{Jiangshan Yu}, {and} \bibinfo{person}{Longxiang Gao}.} \bibinfo{year}{2021}\natexlab{}.
\newblock \showarticletitle{Semantic code search for smart contracts}.
\newblock \bibinfo{journal}{\emph{arXiv preprint arXiv:2111.14139}} (\bibinfo{year}{2021}).
\newblock


\bibitem[Shi et~al\mbox{.}(2022)]%
        {shi2022bytecode}
\bibfield{author}{\bibinfo{person}{Chaochen Shi}, \bibinfo{person}{Yong Xiang}, \bibinfo{person}{Jiangshan Yu}, \bibinfo{person}{Longxiang Gao}, \bibinfo{person}{Keshav Sood}, {and} \bibinfo{person}{Robin Ram~Mohan Doss}.} \bibinfo{year}{2022}\natexlab{}.
\newblock \showarticletitle{A bytecode-based approach for smart contract classification}. In \bibinfo{booktitle}{\emph{2022 IEEE International Conference on Software Analysis, Evolution and Reengineering (SANER)}}. IEEE, \bibinfo{pages}{1046--1054}.
\newblock


\bibitem[Shi et~al\mbox{.}(2023)]%
        {shi2023machine}
\bibfield{author}{\bibinfo{person}{Chaochen Shi}, \bibinfo{person}{Yong Xiang}, \bibinfo{person}{Jiangshan Yu}, \bibinfo{person}{Keshav Sood}, {and} \bibinfo{person}{Longxiang Gao}.} \bibinfo{year}{2023}\natexlab{}.
\newblock \showarticletitle{Machine translation-based fine-grained comments generation for solidity smart contracts}.
\newblock \bibinfo{journal}{\emph{Information and Software Technology}}  \bibinfo{volume}{153} (\bibinfo{year}{2023}), \bibinfo{pages}{107065}.
\newblock


\bibitem[Sridhara et~al\mbox{.}(2010)]%
        {sridhara2010towards}
\bibfield{author}{\bibinfo{person}{Giriprasad Sridhara}, \bibinfo{person}{Emily Hill}, \bibinfo{person}{Divya Muppaneni}, \bibinfo{person}{Lori Pollock}, {and} \bibinfo{person}{K Vijay-Shanker}.} \bibinfo{year}{2010}\natexlab{}.
\newblock \showarticletitle{Towards automatically generating summary comments for java methods}. In \bibinfo{booktitle}{\emph{Proceedings of the IEEE/ACM international conference on Automated software engineering}}. \bibinfo{pages}{43--52}.
\newblock


\bibitem[Suiche(2017)]%
        {suiche2017porosity}
\bibfield{author}{\bibinfo{person}{Matt Suiche}.} \bibinfo{year}{2017}\natexlab{}.
\newblock \showarticletitle{Porosity: A decompiler for blockchain-based smart contracts bytecode}.
\newblock \bibinfo{journal}{\emph{DEF con}} \bibinfo{volume}{25}, \bibinfo{number}{11} (\bibinfo{year}{2017}).
\newblock


\bibitem[Szabo(1996)]%
        {szabo1996smart}
\bibfield{author}{\bibinfo{person}{Nick Szabo}.} \bibinfo{year}{1996}\natexlab{}.
\newblock \showarticletitle{Smart contracts: building blocks for digital markets}.
\newblock \bibinfo{journal}{\emph{EXTROPY: The Journal of Transhumanist Thought,(16)}} \bibinfo{volume}{18}, \bibinfo{number}{2} (\bibinfo{year}{1996}), \bibinfo{pages}{28}.
\newblock


\bibitem[Szabo(1997)]%
        {szabo1997formalizing}
\bibfield{author}{\bibinfo{person}{Nick Szabo}.} \bibinfo{year}{1997}\natexlab{}.
\newblock \showarticletitle{Formalizing and securing relationships on public networks}.
\newblock \bibinfo{journal}{\emph{First monday}} (\bibinfo{year}{1997}).
\newblock


\bibitem[Tu et~al\mbox{.}(2016)]%
        {tu2016modeling}
\bibfield{author}{\bibinfo{person}{Zhaopeng Tu}, \bibinfo{person}{Zhengdong Lu}, \bibinfo{person}{Yang Liu}, \bibinfo{person}{Xiaohua Liu}, {and} \bibinfo{person}{Hang Li}.} \bibinfo{year}{2016}\natexlab{}.
\newblock \showarticletitle{Modeling coverage for neural machine translation}.
\newblock \bibinfo{journal}{\emph{arXiv preprint arXiv:1601.04811}} (\bibinfo{year}{2016}).
\newblock


\bibitem[Vaswani et~al\mbox{.}(2017)]%
        {vaswani2017attention}
\bibfield{author}{\bibinfo{person}{Ashish Vaswani}, \bibinfo{person}{Noam Shazeer}, \bibinfo{person}{Niki Parmar}, \bibinfo{person}{Jakob Uszkoreit}, \bibinfo{person}{Llion Jones}, \bibinfo{person}{Aidan~N Gomez}, \bibinfo{person}{{\L}ukasz Kaiser}, {and} \bibinfo{person}{Illia Polosukhin}.} \bibinfo{year}{2017}\natexlab{}.
\newblock \showarticletitle{Attention is all you need}.
\newblock \bibinfo{journal}{\emph{Advances in neural information processing systems}}  \bibinfo{volume}{30} (\bibinfo{year}{2017}).
\newblock


\bibitem[Wan et~al\mbox{.}(2018)]%
        {wan2018improving}
\bibfield{author}{\bibinfo{person}{Yao Wan}, \bibinfo{person}{Zhou Zhao}, \bibinfo{person}{Min Yang}, \bibinfo{person}{Guandong Xu}, \bibinfo{person}{Haochao Ying}, \bibinfo{person}{Jian Wu}, {and} \bibinfo{person}{Philip~S Yu}.} \bibinfo{year}{2018}\natexlab{}.
\newblock \showarticletitle{Improving automatic source code summarization via deep reinforcement learning}. In \bibinfo{booktitle}{\emph{Proceedings of the 33rd ACM/IEEE International Conference on Automated Software Engineering}}. \bibinfo{pages}{397--407}.
\newblock


\bibitem[Wang et~al\mbox{.}({[n.\,d.]})]%
        {wang2024makes}
\bibfield{author}{\bibinfo{person}{Haoye Wang}, \bibinfo{person}{Zhipeng Gao}, \bibinfo{person}{Tingting Bi}, \bibinfo{person}{John Grundy}, \bibinfo{person}{Xinyu Wang}, \bibinfo{person}{Minghui Wu}, {and} \bibinfo{person}{Xiaohu Yang}.} \bibinfo{year}{[n.\,d.]}\natexlab{}.
\newblock \showarticletitle{What Makes a Good TODO Comment?}
\newblock \bibinfo{journal}{\emph{ACM Transactions on Software Engineering and Methodology}} (\bibinfo{year}{[n.\,d.]}).
\newblock


\bibitem[Wang et~al\mbox{.}(2022)]%
        {wang2022empirical}
\bibfield{author}{\bibinfo{person}{Yilin Wang}, \bibinfo{person}{Xiangping Chen}, \bibinfo{person}{Yuan Huang}, \bibinfo{person}{Hao-Nan Zhu}, {and} \bibinfo{person}{Jing Bian}.} \bibinfo{year}{2022}\natexlab{}.
\newblock \showarticletitle{An empirical study on real bug fixes in smart contracts projects}.
\newblock \bibinfo{journal}{\emph{arXiv preprint arXiv:2210.11990}} (\bibinfo{year}{2022}).
\newblock


\bibitem[Wang et~al\mbox{.}(2021)]%
        {wang2021codet5}
\bibfield{author}{\bibinfo{person}{Yue Wang}, \bibinfo{person}{Weishi Wang}, \bibinfo{person}{Shafiq Joty}, {and} \bibinfo{person}{Steven~CH Hoi}.} \bibinfo{year}{2021}\natexlab{}.
\newblock \showarticletitle{Codet5: Identifier-aware unified pre-trained encoder-decoder models for code understanding and generation}.
\newblock \bibinfo{journal}{\emph{arXiv preprint arXiv:2109.00859}} (\bibinfo{year}{2021}).
\newblock


\bibitem[Wei et~al\mbox{.}(2020)]%
        {wei2020retrieve}
\bibfield{author}{\bibinfo{person}{Bolin Wei}, \bibinfo{person}{Yongmin Li}, \bibinfo{person}{Ge Li}, \bibinfo{person}{Xin Xia}, \bibinfo{person}{David Lo}, {and} \bibinfo{person}{Zhi Jin}.} \bibinfo{year}{2020}\natexlab{}.
\newblock \showarticletitle{Retrieve and Refine: Exemplar-based Neural Comment Generation}. In \bibinfo{booktitle}{\emph{2020 IEEE/ACM International Conference on Automated Software Engineering (ASE)}}. IEEE.
\newblock


\bibitem[Wong et~al\mbox{.}(2015)]%
        {wong2015clocom}
\bibfield{author}{\bibinfo{person}{Edmund Wong}, \bibinfo{person}{Taiyue Liu}, {and} \bibinfo{person}{Lin Tan}.} \bibinfo{year}{2015}\natexlab{}.
\newblock \showarticletitle{Clocom: Mining existing source code for automatic comment generation}. In \bibinfo{booktitle}{\emph{2015 IEEE 22nd International Conference on Software Analysis, Evolution, and Reengineering (SANER)}}. IEEE, \bibinfo{pages}{380--389}.
\newblock


\bibitem[Wong et~al\mbox{.}(2013)]%
        {wong2013autocomment}
\bibfield{author}{\bibinfo{person}{Edmund Wong}, \bibinfo{person}{Jinqiu Yang}, {and} \bibinfo{person}{Lin Tan}.} \bibinfo{year}{2013}\natexlab{}.
\newblock \showarticletitle{Autocomment: Mining question and answer sites for automatic comment generation}. In \bibinfo{booktitle}{\emph{2013 28th IEEE/ACM International Conference on Automated Software Engineering (ASE)}}. IEEE, \bibinfo{pages}{562--567}.
\newblock


\bibitem[Wood et~al\mbox{.}(2014)]%
        {wood2014ethereum}
\bibfield{author}{\bibinfo{person}{Gavin Wood} {et~al\mbox{.}}} \bibinfo{year}{2014}\natexlab{}.
\newblock \showarticletitle{Ethereum: A secure decentralised generalised transaction ledger}.
\newblock \bibinfo{journal}{\emph{Ethereum project yellow paper}} \bibinfo{volume}{151}, \bibinfo{number}{2014} (\bibinfo{year}{2014}), \bibinfo{pages}{1--32}.
\newblock


\bibitem[Xue et~al\mbox{.}(2024)]%
        {xue2024selfpico}
\bibfield{author}{\bibinfo{person}{Zhipeng Xue}, \bibinfo{person}{Zhipeng Gao}, \bibinfo{person}{Shaohua Wang}, \bibinfo{person}{Xing Hu}, \bibinfo{person}{Xin Xia}, {and} \bibinfo{person}{Shanping Li}.} \bibinfo{year}{2024}\natexlab{}.
\newblock \showarticletitle{SelfPiCo: Self-Guided Partial Code Execution with LLMs}. In \bibinfo{booktitle}{\emph{Proceedings of the 33rd ACM SIGSOFT International Symposium on Software Testing and Analysis}}. \bibinfo{pages}{1389--1401}.
\newblock


\bibitem[Yan et~al\mbox{.}(2023)]%
        {yan2023closer}
\bibfield{author}{\bibinfo{person}{Dapeng Yan}, \bibinfo{person}{Zhipeng Gao}, {and} \bibinfo{person}{Zhiming Liu}.} \bibinfo{year}{2023}\natexlab{}.
\newblock \showarticletitle{A Closer Look at Different Difficulty Levels Code Generation Abilities of ChatGPT}. In \bibinfo{booktitle}{\emph{2023 38th IEEE/ACM International Conference on Automated Software Engineering (ASE)}}. IEEE, \bibinfo{pages}{1887--1898}.
\newblock


\bibitem[Yang et~al\mbox{.}(2022)]%
        {yang2022ccgir}
\bibfield{author}{\bibinfo{person}{Guang Yang}, \bibinfo{person}{Ke Liu}, \bibinfo{person}{Xiang Chen}, \bibinfo{person}{Yanlin Zhou}, \bibinfo{person}{Chi Yu}, {and} \bibinfo{person}{Hao Lin}.} \bibinfo{year}{2022}\natexlab{}.
\newblock \showarticletitle{CCGIR: Information retrieval-based code comment generation method for smart contracts}.
\newblock \bibinfo{journal}{\emph{Knowledge-Based Systems}}  \bibinfo{volume}{237} (\bibinfo{year}{2022}), \bibinfo{pages}{107858}.
\newblock


\bibitem[Yin et~al\mbox{.}(2019)]%
        {yin2019codesummarization}
\bibfield{author}{\bibinfo{person}{Pengcheng Yin}, \bibinfo{person}{Graham Neubig}, \bibinfo{person}{Satoshi Sekine}, {and} \bibinfo{person}{Katsuhito Sudoh}.} \bibinfo{year}{2019}\natexlab{}.
\newblock \showarticletitle{Code Summarization with a Semantic Role-based Attention Mechanism}. In \bibinfo{booktitle}{\emph{Proceedings of the 57th Annual Meeting of the Association for Computational Linguistics (ACL)}}.
\newblock


\bibitem[Yu et~al\mbox{.}(2020)]%
        {yu2020smart}
\bibfield{author}{\bibinfo{person}{Xiao~Liang Yu}, \bibinfo{person}{Omar Al-Bataineh}, \bibinfo{person}{David Lo}, {and} \bibinfo{person}{Abhik Roychoudhury}.} \bibinfo{year}{2020}\natexlab{}.
\newblock \showarticletitle{Smart contract repair}.
\newblock \bibinfo{journal}{\emph{ACM Transactions on Software Engineering and Methodology (TOSEM)}} \bibinfo{volume}{29}, \bibinfo{number}{4} (\bibinfo{year}{2020}), \bibinfo{pages}{1--32}.
\newblock


\bibitem[Zhu et~al\mbox{.}(2022)]%
        {zhu2022bytecode}
\bibfield{author}{\bibinfo{person}{Di Zhu}, \bibinfo{person}{Feng Yue}, \bibinfo{person}{Jianmin Pang}, \bibinfo{person}{Xin Zhou}, \bibinfo{person}{Wenjie Han}, {and} \bibinfo{person}{Fudong Liu}.} \bibinfo{year}{2022}\natexlab{}.
\newblock \showarticletitle{Bytecode similarity detection of smart contract across optimization options and compiler versions based on triplet network}.
\newblock \bibinfo{journal}{\emph{Electronics}} \bibinfo{volume}{11}, \bibinfo{number}{4} (\bibinfo{year}{2022}), \bibinfo{pages}{597}.
\newblock


\end{thebibliography}

\end{document}